\documentclass[twocolumn]{aastex631}
\usepackage{gensymb}
\usepackage{amsmath}
\usepackage{float}

\shorttitle{Known pulsars in RACS}
\shortauthors{Anumarlapudi et al.}
\graphicspath{{./}{figures/}}

\begin{document}

\title{Characterizing Pulsars Detected in the Rapid ASKAP Continuum Survey}

\author[0000-0002-8935-9882]{Akash Anumarlapudi}
\affiliation{Center for Gravitation, Cosmology, and Astrophysics,
  Department of Physics, University of Wisconsin-Milwaukee, PO Box
  413, Milwaukee, WI, 53201, USA}
  
\author[0000-0001-9002-6603]{Anna~Ehlke}
\affiliation{Center for Gravitation, Cosmology, and Astrophysics,
  Department of Physics, University of Wisconsin-Milwaukee, PO Box
  413, Milwaukee, WI, 53201, USA}

\author[0000-0001-6607-3710]{Megan~L.~Jones}
\affiliation{Center for Gravitation, Cosmology, and Astrophysics,
  Department of Physics, University of Wisconsin-Milwaukee, PO Box
  413, Milwaukee, WI, 53201, USA}

\author[0000-0001-6295-2881]{David~L.~Kaplan}
\affiliation{Center for Gravitation, Cosmology, and Astrophysics,
  Department of Physics, University of Wisconsin-Milwaukee, PO Box
  413, Milwaukee, WI, 53201, USA}

\author[0000-0003-0699-7019]{Dougal Dobie}
\affiliation{Centre for Astrophysics and Supercomputing, Swinburne University of Technology, Hawthorn, Victoria, Australia}
\affiliation{ARC Centre of Excellence for Gravitational Wave Discovery (OzGrav), Hawthorn, Victoria, Australia}

\author[0000-0002-9994-1593]{Emil Lenc}
\affiliation{CSIRO Space and Astronomy, PO Box 76, Epping, NSW, 1710, Australia}

\author[0000-0002-9415-3766]{James~K.~Leung}
\affiliation{Sydney Institute for Astronomy, School of Physics, University of Sydney, NSW, 2006, Australia}
\affiliation{CSIRO Space and Astronomy, PO Box 76, Epping, NSW, 1710, Australia}
\affiliation{ARC Centre of Excellence for Gravitational Wave Discovery (OzGrav), Hawthorn, Victoria, Australia}

\author[0000-0002-2686-438X]{Tara Murphy}
\affiliation{Sydney Institute for Astronomy, School of Physics, University of Sydney, NSW, 2006, Australia}
\affiliation{ARC Centre of Excellence for Gravitational Wave Discovery (OzGrav), Hawthorn, Victoria, Australia}

\author[0000-0003-1575-5249]{Joshua Pritchard}
\affiliation{Sydney Institute for Astronomy, School of Physics, University of Sydney, NSW, 2006, Australia}
\affiliation{CSIRO Space and Astronomy, PO Box 76, Epping, NSW, 1710, Australia}
\affiliation{ARC Centre of Excellence for Gravitational Wave Discovery (OzGrav), Hawthorn, Victoria, Australia}

\author[0000-0001-8026-5903]{Adam~J.~Stewart}
\affiliation{Sydney Institute for Astronomy, School of Physics, University of Sydney, NSW, 2006, Australia}

\author[0000-0002-9409-3214]{Rahul Sengar}
\affiliation{Center for Gravitation, Cosmology, and Astrophysics,
  Department of Physics, University of Wisconsin-Milwaukee, PO Box
  413, Milwaukee, WI, 53201, USA}

\author[0000-0002-6243-7879]{Craig Anderson}
\affiliation{Research School of Astronomy and Astrophysics, Australian National University, Canberra, ACT 2611, Australia}

\author[0000-0003-4417-5374]{Julie Banfield}
\affiliation{Research School of Astronomy and Astrophysics, Australian National University, Canberra, ACT 2611, Australia}

% \author[0000-0001-6279-4772]{Catherine~L.~Hale}
% \affiliation{CSIRO Space and Astronomy, PO Box 1130, Bentley, WA, 6102, Australia}
% \affiliation{School of Physics and Astronomy, University of Edinburgh, Institute for Astronomy, Royal Observatory Edinburgh, Blackford Hill, Edinburgh EH9 3HJ, UK}

\author[0000-0002-2155-6054]{George Heald}
\affiliation{CSIRO Space and Astronomy, PO Box 1130, Bentley, WA, 6102, Australia}

\author[0000-0001-7464-8801]{Aidan~W.~Hotan}
\affiliation{CSIRO Space and Astronomy, PO Box 1130, Bentley, WA, 6102, Australia}

\author[0000-0002-2819-9977]{David McConnell}
\affiliation{CSIRO Space and Astronomy, PO Box 76, Epping, NSW, 1710, Australia}

\author[0000-0002-3005-9738]{Vanessa A. Moss}
\affiliation{CSIRO Space and Astronomy, PO Box 76, Epping, NSW, 1710, Australia}
\affiliation{Sydney Institute for Astronomy, School of Physics, University of Sydney, NSW, 2006, Australia}

\author{Wasim Raja}
\affiliation{CSIRO Space and Astronomy, PO Box 76, Epping, NSW, 1710, Australia}

\author[0000-0003-1160-2077]{Matthew~T.~Whiting}
\affiliation{CSIRO Space and Astronomy, PO Box 76, Epping, NSW, 1710, Australia}

\begin{abstract}
We present the detection of 661 known pulsars observed with the Australian SKA Pathfinder (ASKAP) telescope at 888\,MHz as a part of the Rapid ASKAP Continuum Survey (RACS). Detections were made through astrometric coincidence and we estimate the false alarm rate of our sample to be $\sim0.5$\%. Using  archival data at 400 and 1400\,MHz, we estimate the power law spectral indices for the pulsars in our sample and find that the mean spectral index is $-1.78\pm 0.6$. However, we also find that a single power law is inadequate to model all the observed spectra. With the addition of the flux densities between 150\,MHz and 3\,GHz from various imaging surveys, we find that up to 40\% of our sample shows deviations from a simple power law model. Using Stokes V measurements from the RACS data, we measured the circular polarization fraction for 9\% of our sample and find that the mean polarization fraction is $\sim 10$\% (consistent between detections and upper limits). Using the dispersion measure (DM) derived distance we estimate the \textit{pseudo luminosity} of the pulsars and do not find any strong evidence for a correlation with the pulsars' intrinsic properties.
\end{abstract}


\section{Introduction} \label{sec:intro}

Because of the complexity involved in modeling a pulsar's magnetosphere, a complete theory of pulsar emission that explains the diverse observed emission properties remains to be understood \citep{Goldreich1969, Sturrock1971, Ruderman1975, Krause-Polstorff1985, Taylor1986, Cerutti2017}. 
Observational evidence in this regard provides a very useful avenue in phenomenologically understanding the underlying emission process\citep[e.g.,][]{Radhakrishnan1969}. 

Spectral and polarimetric signatures of the observed emission are two of the most common properties that can be measured in a large number of pulsars and hence can provide clues about the pulsar's emission mechanism. The observed spectrum in pulsars is usually characterized by a steep power law, $S_\nu = S_0\  \nu^{\alpha}$, typically with power law index $\alpha<-1$ \citep{Bates2013}. In addition, pulsars are one of the small number of object classes in which the emission can be highly polarized, both linearly and circularly \citep{handbook}. Combing the spectral and polarimetric properties of the pulsars can hence provide an alternate way to the routine periodic searches that are used to discover pulsars --- through imaging techniques that are independent of the pulsed emission (e.g., \citealt{1982Natur.300..615B,Navarro1995,Crawford00, Dai18,kaplan2019, WangY2022}).

Traditionally, most pulsars are discovered through periodicity searching techniques, where the signal is dedispersed and then searched for periodicities. Later follow-up observations then add up the emission from individual pulses in phase.  Average properties like flux densities can be difficult to measure reliably from such observations as they rely on accurate knowledge of the telescope gain, sky background temperature, pulse duty cycle, and more \citep{handbook}. In contrast, continuum emission from interferometric images provides a useful alternative to discovering and characterizing  pulsars. \cite{Navarro1995} used  imaging techniques to find a steep-spectrum, highly polarized source, that revealed a 2.3\,ms pulsar, PSR J0218+4232, in follow-up periodic searches in which a significant amount of radio energy is not pulsed. Similarly \cite{WangY2022} discovered a circularly polarized transient, which revealed a $\sim$322~ms pulsar, PSR J0523$-$7125, in the Large Magellanic Cloud (LMC). Follow-up observations showed it was brighter than all previously discovered pulsars in the LMC but might have been missed in the blind periodic searches because of its large pulse width and steep spectrum.  

With the advent of all-sky radio imaging surveys, studies searching for pulsars through imaging are re-discovering an increasing number of pulsars \citep[e.g., and in some cases serendipitously discovering new pulsars;][]{Navarro1995, Kaplan1998, Han1999, Kouwenhoven2000, tgss, Bhakta2017, WangY2022}. In addition, imaging surveys can be extremely fruitful in identifying transients that show unusual polarization properties and/or variability \citep{kaplan2019,Wang2022} and hence studying the spectral and polarimetric properties of these transients can be used to identify/discover pulsars that show variability through scintillation \citep{Crawford00,Dai16,Dai17}. Finally, imaging surveys can measure properties like flux densities reliably for many objects \citep[e.g.,][]{Bell16,MWA2017}, characterizing spectral properties and variability.
%Such a methodology is used by \cite{Wang2021} to discover a highly polarized Galactic center radio transient (GCRT). 
 %Although stars are usually very faint at radio wavelengths, stars with high magnetic activity can emit bursts that can be identified through variability. Using the data from Australian Square Kilometre Array Pathfinder (ASKAP), \cite{Joshua2021} studied the circular polarization in stars demonstrating the ability to look for polarized sources in imaging surveys. 
%In this paper, we focus on the pulsar population and present the detections of known pulsars in the Rapid ASKAP Continuum Survey (RACS) data.

Although the observed spectrum is usually modeled by a power law,  the exact value of power law index $\alpha$ is not very well determined; Using a sample of 280 pulsars observed at 408, 606, 925, 1408 and 1606\,MHz \cite{Lorimer1995} found $\alpha=-1.6\pm0.3$. Analyzing the same data set, but extended to include higher and lower frequencies \cite{Maron2000} found $\alpha=-1.8\pm0.2$. \cite{Bates2013} tried to remove the observational biases to predict the intrinsic pulsar spectrum and found $\alpha=-1.4\pm1.0$. With a sample of 441 pulsars observed at 728, 1382, and 3100\,MHz and \cite{jankowski18} finds $\alpha=-1.6\pm0.54$. All of the flux density measurements from the above studies were derived using single-dish observations. Moreover, there are cases where a simple power law is not adequate to completely describe the spectrum; most common are the low and the high-frequency deviations of the spectrum \citep[][estimate that at most $\sim10$\% of their sample show the preference for more complex models]{Maron2000}. Using flux densities derived through imaging survey in 60 pulsars and combing archival data \cite{MWA2017} found that a single power law is inadequate to fit the observed variation in as much as 50\% of their sample with a broken power providing a better fit, although their data, taken at $\sim$200\,MHz is more sensitive to pulsars at lower flux densities and hence more sensitive to low-frequency variations. However, none of the above studies find any obvious sub-population that prefers a broken power law fit. In addition to this is the question of whether the spectral index is consistent between the normal and ``recycled'' pulsars -- \cite{Kramer1998} found no evidence for such a disparity between the populations with the spectral index being consistent. However, \citet{tgss} found that the spectra of millisecond pulsars are steeper than the normal pulsars, although they attribute this to their survey's selective preference of being sensitive to pulsars at a lower frequency (150\,MHz). %In this regard, RACS provides useful flux measurements at 888\,MHz which can be used with flux measurements at lower and higher frequencies to independently estimate the distribution of spectral indices.

%Pulsar emission, in general, is observed to be polarized, showing evidence for both linear and circular polarizations. 
Similarly, studies done so far on polarimetric measurements measure both linear and circular polarizations. An initial study done by \citet{Gould1998} finds that the linear polarization is $\sim 20-40$\% and the circular polarization is $\sim 8$\%, but with a high degree of scatter, with individual pulsars capable of showing 100\% linear polarization. Using a sample of 24 millisecond pulsars (MSPs) observed at 730, 1400, 3100\,MHz, \cite{Dai2015} found that the level of circular polarization is $\sim 8-10$\% across the three frequencies. Similarly \cite{Johnson2018} used a sample of 600 pulsars observed at 1.4\,GHz to find a mean circular polarization $\sim9$\%. Using a sample of 40 normal pulsars, \cite{Sobey2021} found that the circular polarisation changes between $\sim$ 960 and 3820 MHz roughly by 4\% with a mean polarization of $\sim 16$\%. A more recent study \citep{Oswald2023} finds a consistent circular polarization fraction (on a 5\% level). \cite{Xilouris1998} found that the evolution of polarization fraction with frequency is more complex in milli-second pulsars than the normal pulsars, which makes it interesting to study the frequency dependence of the polarization fraction.

In this paper, we present the results of a search for detected radio pulsars using the Australian SKA Pathfinder telescope (ASKAP; \citealt{askap}), an interferometric array of 36 dish antennas, each 12\,m in diameter, achieving a resolution of 15\arcsec.  We make use of the total intensity (Stokes I) and circular polarization (Stokes V) sky maps and source catalogs to detect and characterize the radio emission from pulsars. This paper is organized as follows: Section~\ref{sec:data_ana} describes the data reduction and the pulsar sample selection methodology. In Section~\ref{sec:results}, we present the source properties; astrometric, spectral characterization, and polarization measurements of the pulsars in our sample. In Section~\ref{sec:discussion}, we provide the implications of our results, combining them with the findings of past studies before concluding in Section~\ref{sec:conclusion}. 

\section{Data Analysis} \label{sec:data_ana}

\subsection{Data reduction}\label{sec:data_red}

Data were collected as a part of the Rapid ASKAP Continuum Survey (RACS) survey \citep{racs}, an all-sky survey (south of declination $+51\degree$) initially observing at 888~MHz with a bandwidth of 288~MHz. The observations were carried out from 2019 April 21 through 2020 June 21 (constituting the first RACS data set; RACS DR 1) \citep{racs} and cover the sky south of $+41\degree$ declination with an integration time of $15$\,min which were used to generate both Stokes I and Stokes V images. Data were processed using the ASKAP\textsc{soft} package \citep{askapsoft}, which includes methods for flagging, calibration, and generating images for each primary beam. Beams were then linearly mosaicked to generate a single image for individual tiles. Flux density calibrations were done using PKS~B1934--638, which is the primary reference source used for ASKAP \citep{askap}. A more detailed description of reduction techniques used for RACS data can be found in \cite{racs}.

\subsection{Sample Selection} \label{subsec:sample_selection}

We selected all the pulsars from the ATNF catalog \citep[][v1.69]{psrcat} that are in the RACS DR1 footprint (declination $<41$\degree). At the time of writing, this constituted a sample of 3122 pulsars. In order to avoid source confusion, we removed all the pulsars that are known to be associated with globular clusters, resulting in 2915 pulsars. Sources with astrometric positional errors larger than 10\arcsec\, were removed to avoid association with background sources in RACS purely by chance ending up with a sample of 2235 pulsars ($\sim 71$\% of the original sample). Pulsars can have significant proper motion and hence can result in positional mismatches if not accounted for (e.g., J1856+0912, the pulsar with the highest proper motion in our sample, $331.2\,{\rm mas\,yr}^{-1}$, can have an apparent motion of $\sim$4.2\,\arcsec\, between the reference epoch and the RACS epoch). Hence, we corrected for the proper motion of the pulsars whenever available to estimate the pulsars' positions at the RACS epoch. 
We determined the search radius around a RACS source such that the probability of finding a source with a positional offset $r$ due to the uncertainty $\sigma_{\rm RACS}$ is greater than the probability of finding the closest neighboring source at the same offset given the local background density\footnote{Using a patch of radius 1\degree.} $\rho_{\rm RACS}$ of the RACS survey.
As described by \cite{racs}, there can be systematic uncertainty of $\sim$2\arcsec\, for sources in the RACS survey and hence choose a conservative error of 2\arcsec\, on all of the RACS sources. For RACS DR1, we find this search radius to be $\sim$10\arcsec. 

We selected all the RACS sources whose positions are consistent with the pulsar's position to within 10\arcsec\ taking into account the uncertainties in both the RACS and the pulsar's astrometric measurements. This resulted in 661 matches: 600 with only a Stokes I match, and 61 sources with both  Stokes I and Stokes V matches. We visually inspected all 661 matches manually to look for source confusion in the presence of multiple close-by sources but all of the sources seemed reasonable detections. \footnote{As part of a search of circularly polarized sources in the Variable and Slow Transients (VAST; \citealt{Murphy2021}) survey (Pritchard et al., in prep.), we identified a source in the vicinity of the pulsar B1353$-$62.  Follow-up observations with the Ultra-Wide Low \citep{hobbs2020} receiver on the  
Parkes 64m ``Murriyang" radio-telescope determined that the polarized source was in fact B1353$-$62, with pulsations visible from 700\,MHz to 4\,GHz.  We therefore update the position of the pulsar to be RA: $13^{\rm h}56^{\rm m}55.4^{\rm s}$ and DEC: $-62\arcdeg 30\arcmin 07\arcsec$ with an uncertainty (including systematic) of 2.5\arcsec\ in either direction, and include it in this analysis.} Taking the local density around these sources, we estimate that for a 10\arcsec\, search radius, there can be at most 4 false positives in Stokes I matches and $\ll1$ Stokes V match with 95\% confidence that could have been identified by chance and hence the false alarm rate for our sample is $\sim$0.5\%.

\section{Results}\label{sec:results}
\subsection{Source properties}
\begin{figure}
    \plotone{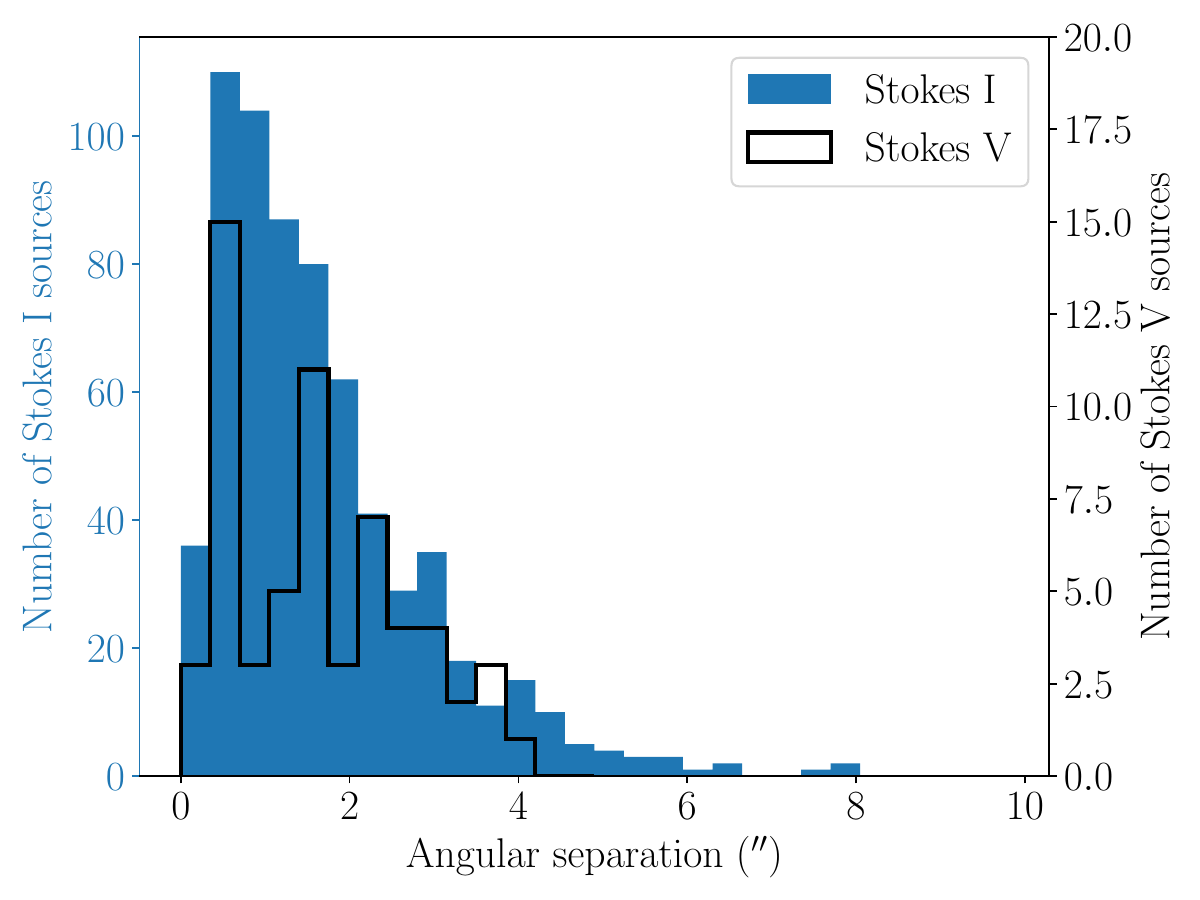}
    \caption{Distribution of spatial offset between the  ATNF catalog positions and the  RACS positions for the pulsar cross matches. The blue histograms show the distribution for RACS Stokes I cross matches and the black histogram shows the same for RACS Stokes V sources. 
    %The median separation for Stokes I sources is 1\farcs3$\pm$1\farcs1 and for Stokes V sources is 1\farcs5$\pm$1\farcs0, consistent within 1-$\sigma$ of each other.
    }
    \label{fig:sep_dist}
\end{figure}

The distribution of sky positions of all the pulsar cross matches in the RACS data is shown in Figure~\ref{Fig:skypos} (sample detection images are shown in Appendix~\ref{sec:app}). All the pulsars from the ATNF catalog are shown in gray dots, while the Stokes I detections are shown in orange diamonds, and simultaneous Stokes I and Stokes V detections are shown as blue stars. We see that most of the detections lie along the Galactic plane (shown in the green stripe), tracing the Galactic pulsar population.
Spatial offsets were calculated between the positions of the RACS detections and the ATNF catalog positions, and the resulting distribution is shown in Figure~\ref{fig:sep_dist}. We find that $\sim$98\% of Stokes I detections and 100\% of the Stokes V detections are within 5\arcsec\ of the pulsar's position suggesting that most of the candidates are likely the pulsar cross matches as opposed to the random uniform distribution expected for background noise. We find that the median separation between the RACS source and the ATNF position for Stokes I detection is 1\farcs3$\pm$1\farcs1 and for Stokes V detection is 1\farcs5$\pm$1\farcs0 (consistent to within 1-$\sigma$ of each other).

\begin{figure*}[!htb]
   \plotone{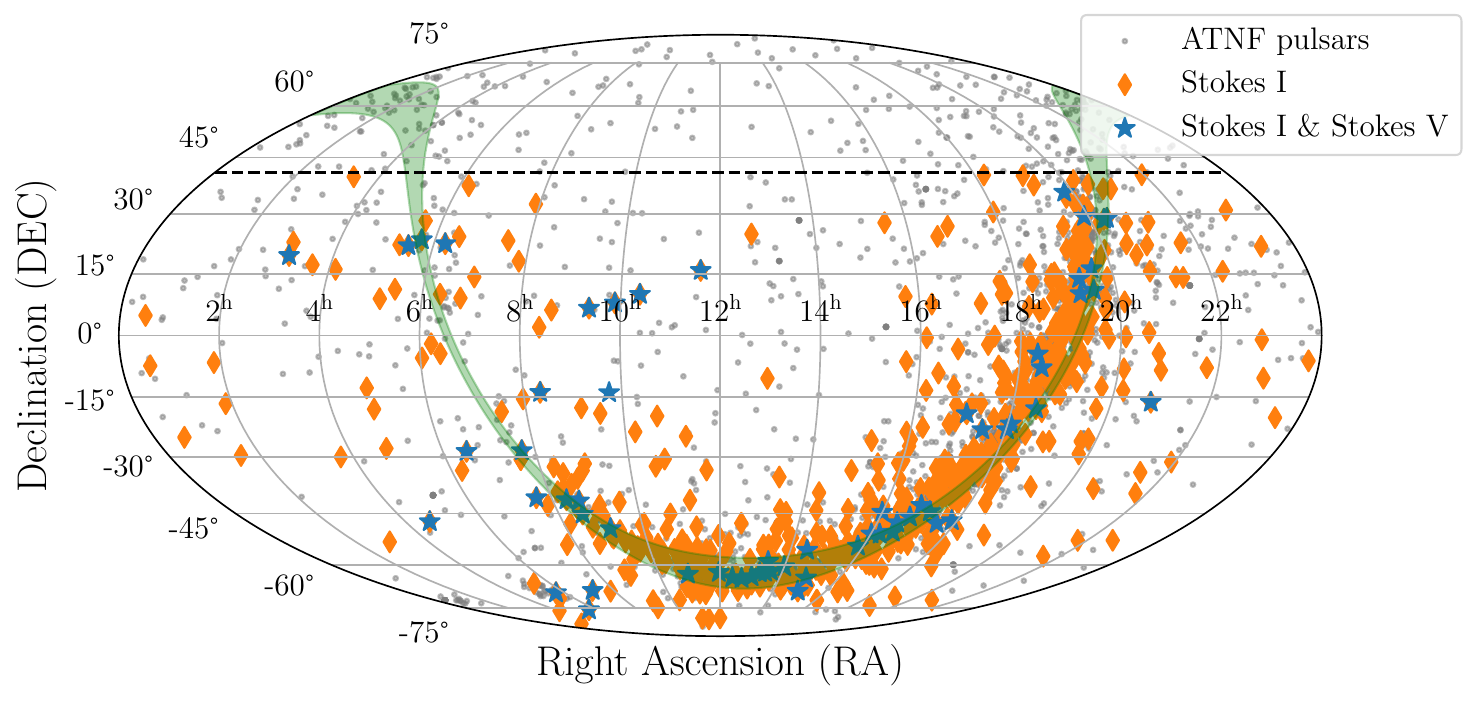}
    \caption{The spatial distribution of all the RACS sources that have an ATNF pulsar within 10\arcsec. Sky positions (in mollweide projection) of all the ATNF pulsars are shown as gray dots, the RACS Stokes I pulsar detections are shown as orange diamonds, and simultaneous RACS Stokes I and Stokes V detections are shown as blue stars. The black dashed line shows the declination limit, +41\degree, for the RACS data release 1. The Galactic plane within  $|b|<5\degree$ is shown in the green shaded region.}
    \label{Fig:skypos}
\end{figure*}

\subsection{Completeness}\label{sec:completeness}

\begin{figure*}
    \centering
    \gridline{
    \fig{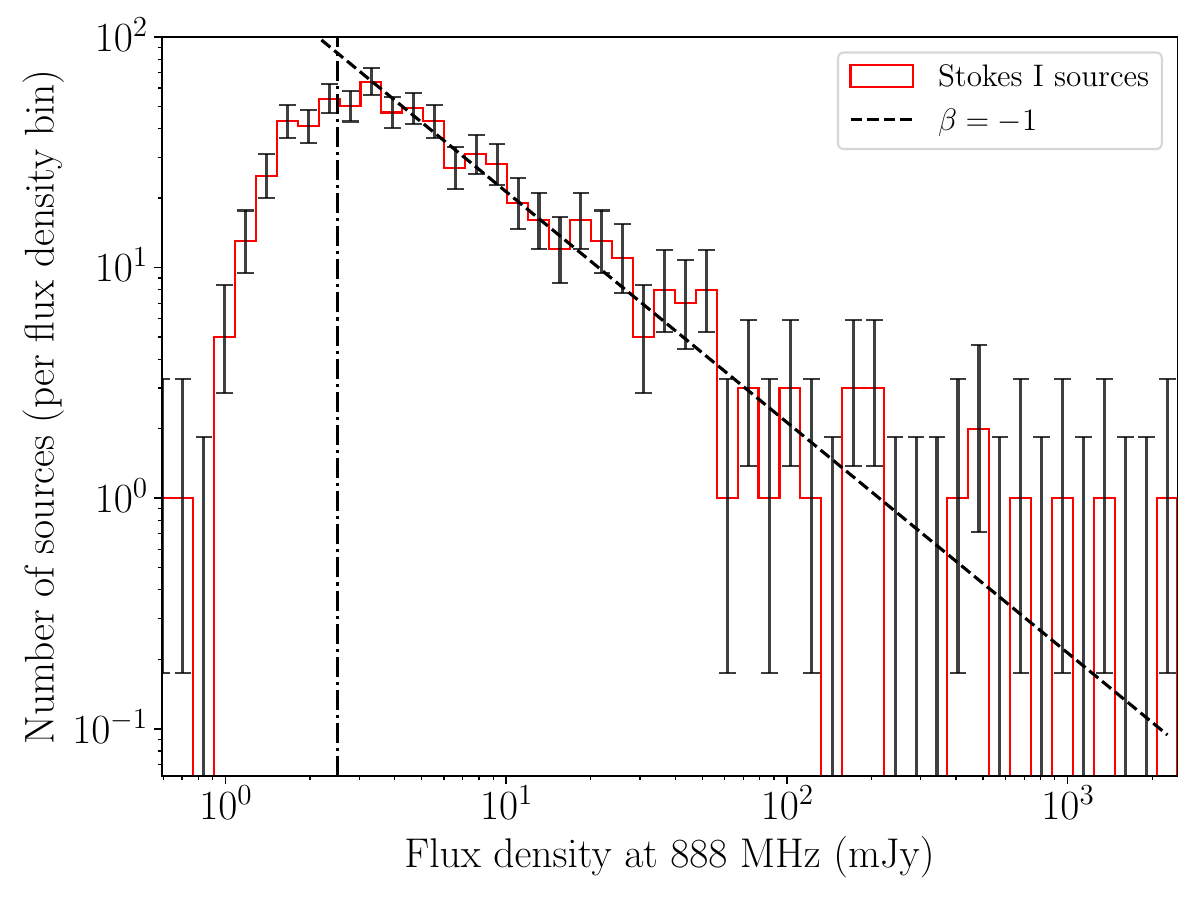}{0.45\textwidth}{(a) Number density of Stokes I sources}
    \fig{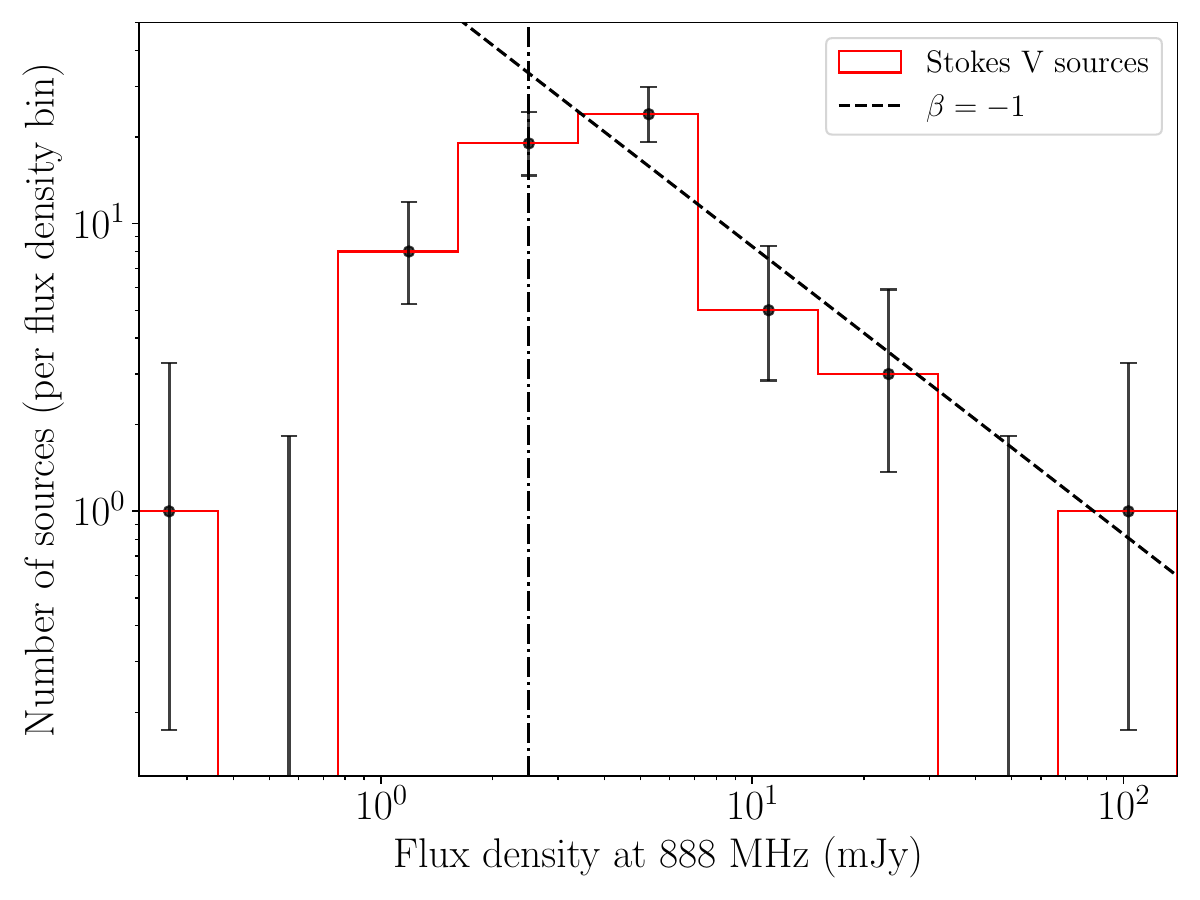}{0.45\textwidth}{(b) Number density of Stokes V sources}
    }
    \caption{The distribution of the number of pulsars as a function of flux density detected in Stokes I images (\textit{Left}) vs Stokes V images (\textit{Right}) is shown by the red histogram. 
    %We detected a handful of very bright ($>$ Jy) sources with most of the pulsars . The median value of flux for Stokes I sources is 4.23 mJy/beam and for Stokes V sources is 3.45 mJy/beam. 
    Shown in the dashed black line is a power law fit to the number of sources per flux density bin assuming a two-dimensional (Galactic) distribution of standard candles ($\beta=-$1). The vertical black dashed-dotted line shows the flux density limit below which we see large deviations in the observed number of sources, suggesting that the survey is complete above $\approx 2\,$mJy.}
    \label{fig:flux_dist}
\end{figure*}

The distribution of pulsars detected in the RACS survey as a function of their flux density is shown in Figure~\ref{fig:flux_dist}. We see that most of the pulsars in the sample have flux densities of a few mJy (toward the detection limit) with a handful of them detected at very high flux densities ($>$ Jy). The red histogram shows the observed number of the sources per flux density bin and the black error bars show the asymmetric 1-$\sigma$ upper and lower limits (calculated according to \citealt{gehrels1986}). For a uniform spatial distribution of standard candles, the number of observable sources with the flux density $>S_{\nu}$ follows a simple power law, $N(>S_\nu) = A S_{\nu}^{\beta} $, where $\beta$ is the power law index. For a two-dimensional distribution of sources in the Galactic plane (pulsars have a typical scale height of $\sim$300-350\,pc, which is much smaller  than their distances $\sim$2-6\,kpc; \citealt{Mdzinarishvili2004,Lorimer2006}) $\beta=-1$ and hence we fit the observed number of pulsars with a power law of slope $-1$. Below a certain flux density limit, we will see a drop-off from the expected distribution which can be used to assess the (in-)completeness of the survey. The black dashed line in Figure~\ref{fig:flux_dist} shows the best fit for the number density of sources assuming $\beta=-1$. 

We can see that for Stokes I (\textit{left}), there seems to be a turnover at $\sim$2\,mJy (marked by the black dashed-dotted line) below which we see a rapid drop in the number of detected sources suggesting that the survey is complete above a flux density level of $\sim$2\,mJy. A similar analysis for Stokes V sources is difficult due to the small number of sources per bin, but the completeness limit estimated for Stokes I matches seems to be consistent with the Stokes V population. This is higher than expectations based on the noise in the RACS images, roughly $0.25\,{\rm mJy\,beam}^{-1}$ at high latitudes, leading to a $1.25\,$mJy limit at 5-$\sigma$, but reasonable when the locations of the pulsars in the Galactic plane (with higher confusion noise) are considered.

We then compared the astrometric and spin properties of pulsars detected in RACS with the overall population of pulsars from the ATNF catalog using a non-parametric test, the Anderson-Darling (AD) test \citep{andersondarling,andersondarling_ksamp}
% the Kolmogorov–Smirnov (KS) test \citep{Kolmogorov, Smirov1948}

\begin{figure}
    \centering
    \plotone{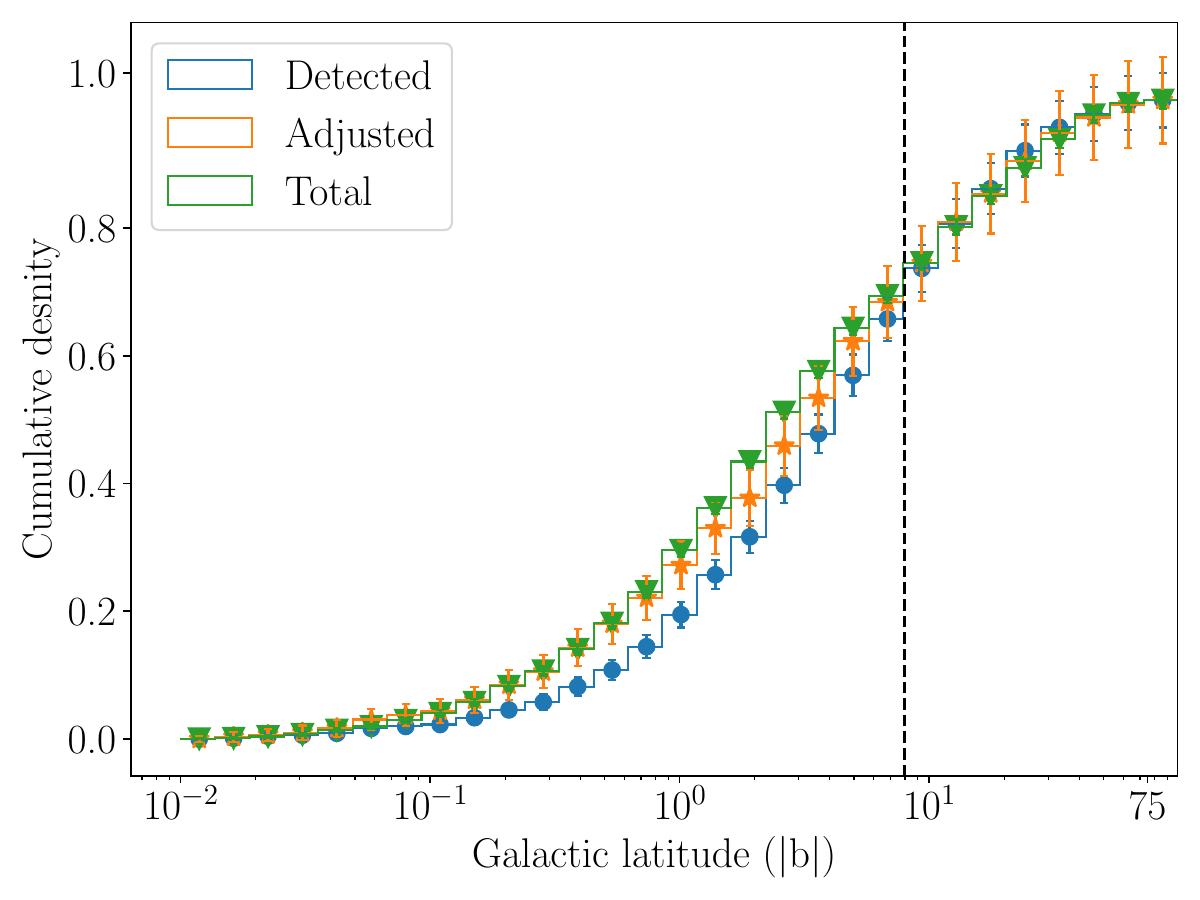}
    \caption{The distribution of Galactic latitudes of the pulsars detected in our sample. The blue histogram shows the Galactic latitudes of the pulsars detected in the RACS survey and the green histogram shows the overall distribution of the pulsars from the ATNF catalog. We see that there is a deficit in the number of pulsars detected for $|b|<8\degree$ (vertical dashed line). We then remove all the detected pulsars below 3\,mJy, and the resulting distribution (orange histogram)  traces the expected population.}
    \label{fig:galb_bias}
\end{figure}

Figure~\ref{fig:galb_bias} shows a comparison of the Galactic latitudes of the detected population with the overall population of pulsars. We observe a deficit in the number of pulsars detected for pulsars that lie close to the Galactic plane ($|b|<8\degree$). The AD-test yields a p-value of 0.004, which provides evidence against the null hypothesis that the population recovered from our survey and the population from the periodic searches are similar. This can be expected because the background noise is higher for sources closer to the Galactic plane, reducing the number of detections. 
%To test this, we estimated the number of pulsars that would be detectable if we lowered the flux density limit of the survey from 2\,mJy to 1.5\,mJy (see Figure~\ref{fig:flux_dist}) and distributed them randomly within the Galactic plane. 
To test this we estimated the number of pulsars that would be detected if the flux density limit for the detection were higher (3\,mJy). This would remove all the fainter sources that were detected at the higher latitudes because of the lower background noise compared to the ones in the plane.
The orange histogram in Figure~\ref{fig:galb_bias} shows this expected number and we see that it traces the overall observed population from the periodic searches (p-value of 0.25 for the AD-test).  We conclude that the low-latitude deficit that we observe in our data is attributable to the increased background noise in the Galactic plane that causes the sources with lower flux densities to be preferentially detected at higher latitudes. We repeated a similar exercise for the spin period distributions but we do not find any evidence against the null hypothesis, so we conclude that the detection of sources in imaging surveys like ASKAP is not dependent on the spin period (as expected).

%around which results in the best fit value of spectral index to be $\alpha=-1.55\pm0.03$ for Stokes I sources and $\alpha=-1.15\pm0.15$ for Stokes V sources. Detailed information on these detections is provided in Table~\ref{tbl-fluxinfo}.

% J0613+3731 --- LOFAR detected at and timed it to <1'' accuracy. Taking RACS uncertainity as 2.5'', the separation is 14'', so purely by chance coincidence? J0622+3749 (https://arxiv.org/pdf/2106.09396.pdf reports an extended gamma-ray emission)
% J0942 seems to be a vary faint source in the close proximity of a bright source.
% J1015 seems to be sitting on a background star/galaxy
% J1530 seems a bit extended but the sep is 1.73'', so may be a match
% J1747-2958 has a wake but that wake was detected even in discovery

\subsection{Flux Density Uncertainties Due To Scintillation}\label{sec:scint}
In addition to the statistical uncertainties in the flux densities due to measurement noise, there can be additional  uncertainties in the flux density due to diffractive scintillation.
Inhomogeneities in the ionized interstellar medium (ISM) cause random perturbations in the phases of the radio signals which can interfere to produce a scintillation pattern at the receiver. Hence the observed flux density can be strongly modulated if the scintillation is extreme. The strength of scintillation (characterized by the number of brightness maxima in the time-frequency plane, known as ``scintles"), can be described by the diffractive scintillation bandwidth ($\Delta \nu_d$) and diffractive scintillation timescale ($\Delta t_d$) \citep[see][for a review]{cordes1991}. The number of scintles in the frequency and time domain are given by
\begin{eqnarray}
N_{\nu} &=& 1+ \eta \frac{\Delta \nu}{\Delta \nu_d}\nonumber \\
N_{t} &=& 1+ \eta \frac{\Delta t}{\Delta t_d}\nonumber
\end{eqnarray}

\noindent where $\Delta \nu$ and $\Delta t$ are the observing bandwidth and the observational duration (for RACS observations, these are 288\,MHz and $\sim$1000\,s respectively) and $\eta \sim 0.1-0.2$ (we considered $\eta=0.15$). For diffractive scintillation, the fractional error in flux density $\Delta S_\nu/S_\nu \sim (N_{\nu} N_t)^{-0.5}$. 

\begin{figure}
    \plotone{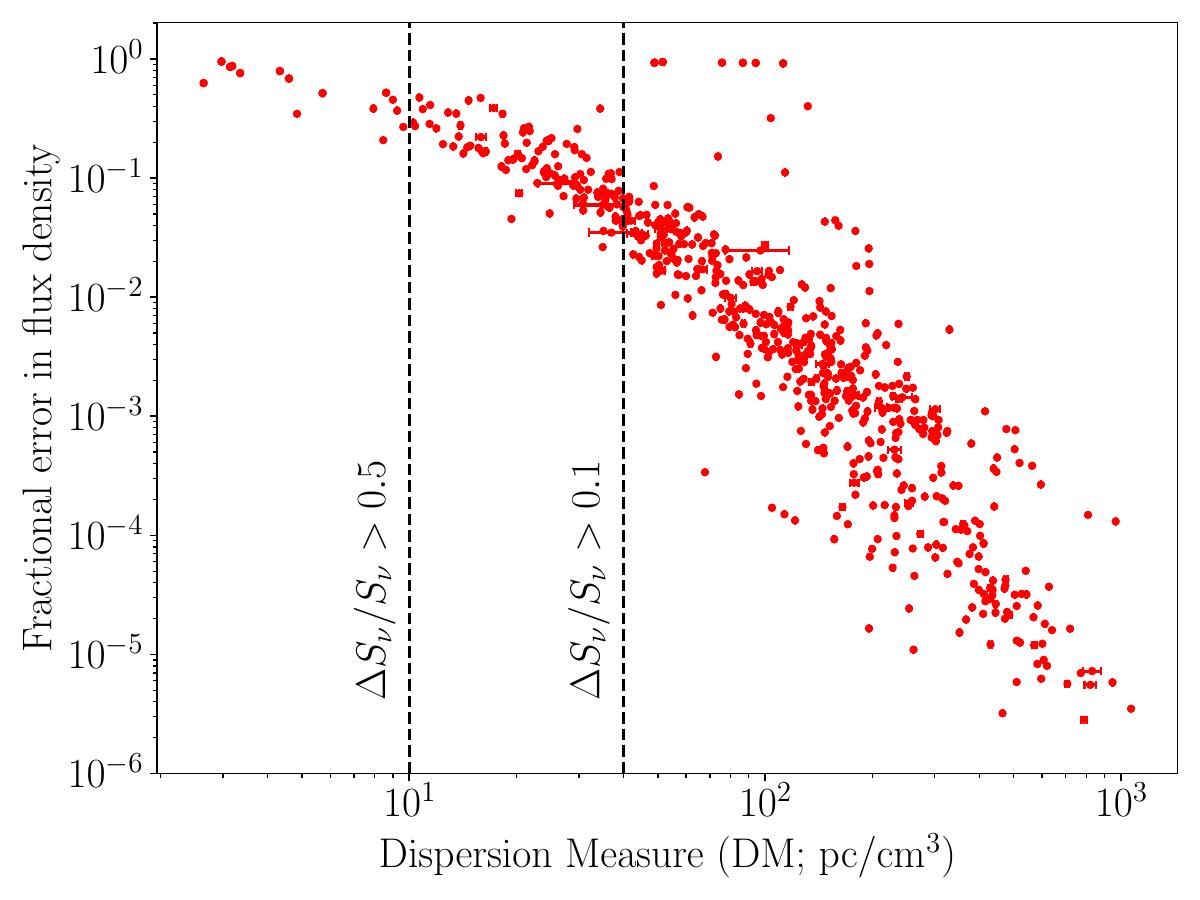}
    \caption{For pulsars identified in RACS data, the fractional error in flux density due to diffractive scintillation as a function of the dispersion measure is shown as the red dots.  
    % Shown as the red dots are the pulsar in which the fraction error is $>0.5$, in violet are the pulsars in which the error is $>0.1$, and in yellow is the rest of the sample. 
    We see that only a few pulsars (18 pulsars in the sample --- $\sim$2.5\% of the sample) are strongly affected (having a fractional error $>0.5$) by scintillation and most of them have DM $<$ 10\,pc/cm$^3$. We can also see that most of the pulsars that have a fractional error of $>0.1$ have DM $<$ 40\,pc/cm$^3$.
    }
    \label{fig:scint_dist}
\end{figure}

Figure~\ref{fig:scint_dist} shows the fractional error in the flux density that can be caused due to diffractive scintillation as a function of the dispersion measure (DM) for all the RACS sources in our sample. We see that most of the pulsars in our sample are in the regime where the errors due to diffractive scintillation are not very significant (fractional error $<0.5$), but there are a few pulsars ($\sim 2.5\%$ of the entire sample), that have a fractional error of $>$0.5. This limit on fractional error can be roughly translated to a limit on DM --- most of the pulsars with fractional error $>0.5$ have DM $<$ 10\,pc/cm$^3$ and with fractional error $>0.1$ have DM $<$ 40\,pc/cm$^3$.

In addition to diffractive scintillation, pulsars are also known to suffer long-term intensity variations caused by the large-scale structures in the ISM due to refractive interstellar scintillation (RISS; \citealt{Seiber1982}). This can cause the flux density to vary over days to months which can be a limiting factor when modeling the pulsar spectra using non-simultaneous flux density measurements (see \S\ref{sec:spec}). Following \cite{Romani1986, Bhat1999} we estimate the fractional error in the flux density due to RISS for the pulsars in our sample. We use the \cite{NE2001} electron density map to estimate the distance to the pulsar and the scattering measure (see \citealt{Kaplan1998, Bhat1999}). We find that for most of the pulsars in our sample the fractional error varies from 6\% -- 18\% (16$^{\rm th}$ and  84$^{\rm th}$ percentiles).

\subsection{Spectral Index Distribution}\label{sec:spec}
\begin{figure}
    \plotone{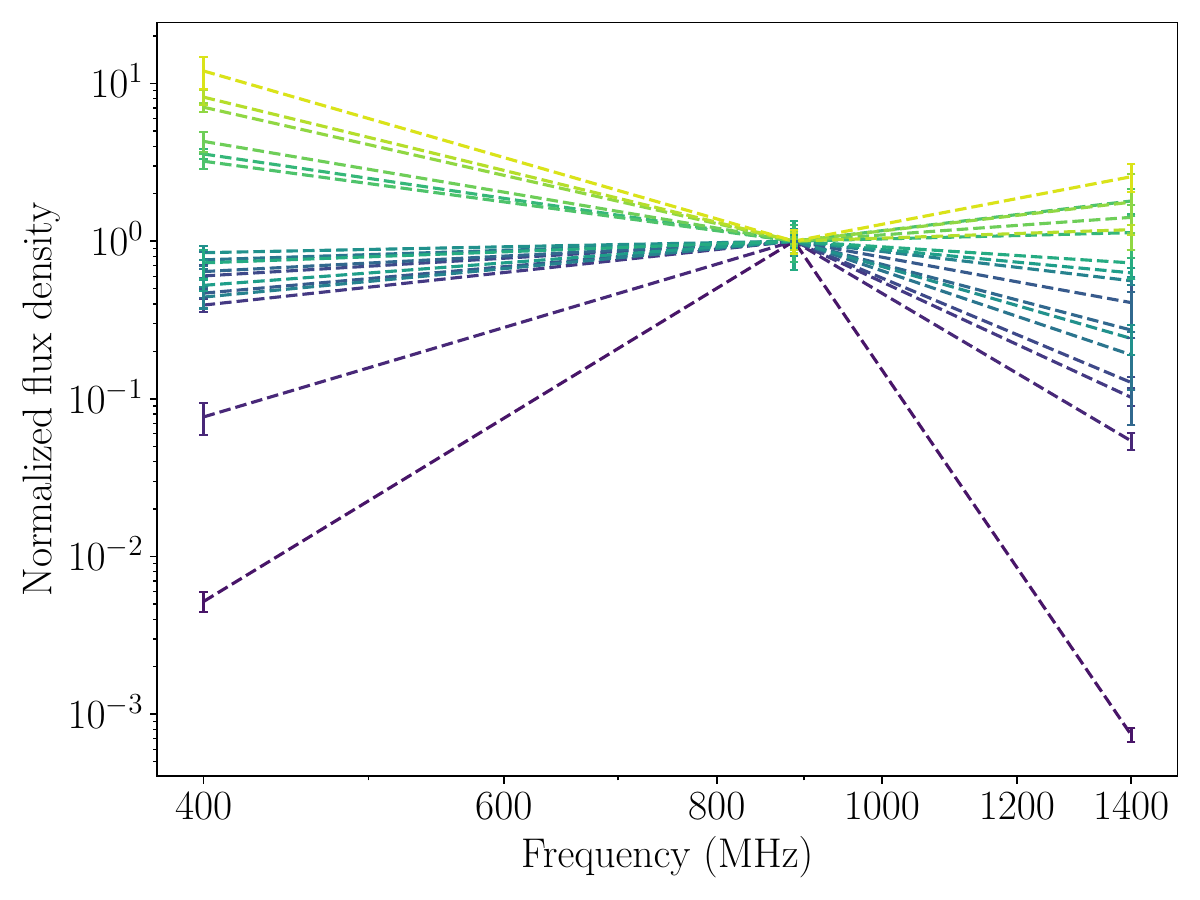}
    \caption{The sample of 18 pulsars that deviate significantly from monotonic spectral energy distributions. Each line in the plot shows the spectrum for a single pulsar using the flux density measurements taken at 400, 888, and 1400\,MHz. The color scheme represents the extent of deviation from a flat spectrum (violet/yellow corresponds to maximum deficit/excess) added over the three frequencies. The flux density at 888\,MHz is normalized to 1. We excluded such pulsars from the fit when using a simple power law to model the spectrum.}
    \label{fig:non_mono_pulsars}
\end{figure}

From the sample of 661 pulsars that had a RACS counterpart within 10\arcsec, 168 pulsars had measured flux densities at 400\,MHz and 1400\,MHz\footnote{We note that these measurements come from a variety of sources and may have mixed reliability.} (ATNF catalog). Table~\ref{tab:flux_tab} gives the flux density measurements for the 168 pulsar sample. We performed a least-squares fit to find the spectral index, assuming a power law distribution, using the flux densities at 400\,MHz (ATNF catalog), 888\,MHz (RACS low DR1), and 1400\,MHz (ATNF catalog). From a visual examination, we excluded 18 pulsars where the flux densities can not be modeled by a single power law since they show non-monotonic behavior, either from a more complex spectral behavior \citep{Bates2013, Swainston2022} or from variability among non-simultaneous measurements. Figure~\ref{fig:non_mono_pulsars} shows the 18 pulsars in our sample that show non-monotonic spectral evolution and hence can not be described by a power law. For the rest of the sample, we find that not all the pulsars can be adequately modeled by a power-law spectrum; out of the 150 pulsars that show monotonic spectral variation, only 90 pulsars (60\% of the sample) can be well modeled by a power law (they have a goodness of the fit, reduced $\chi^2$ $\leq2$). 

\begin{figure}
    \plotone{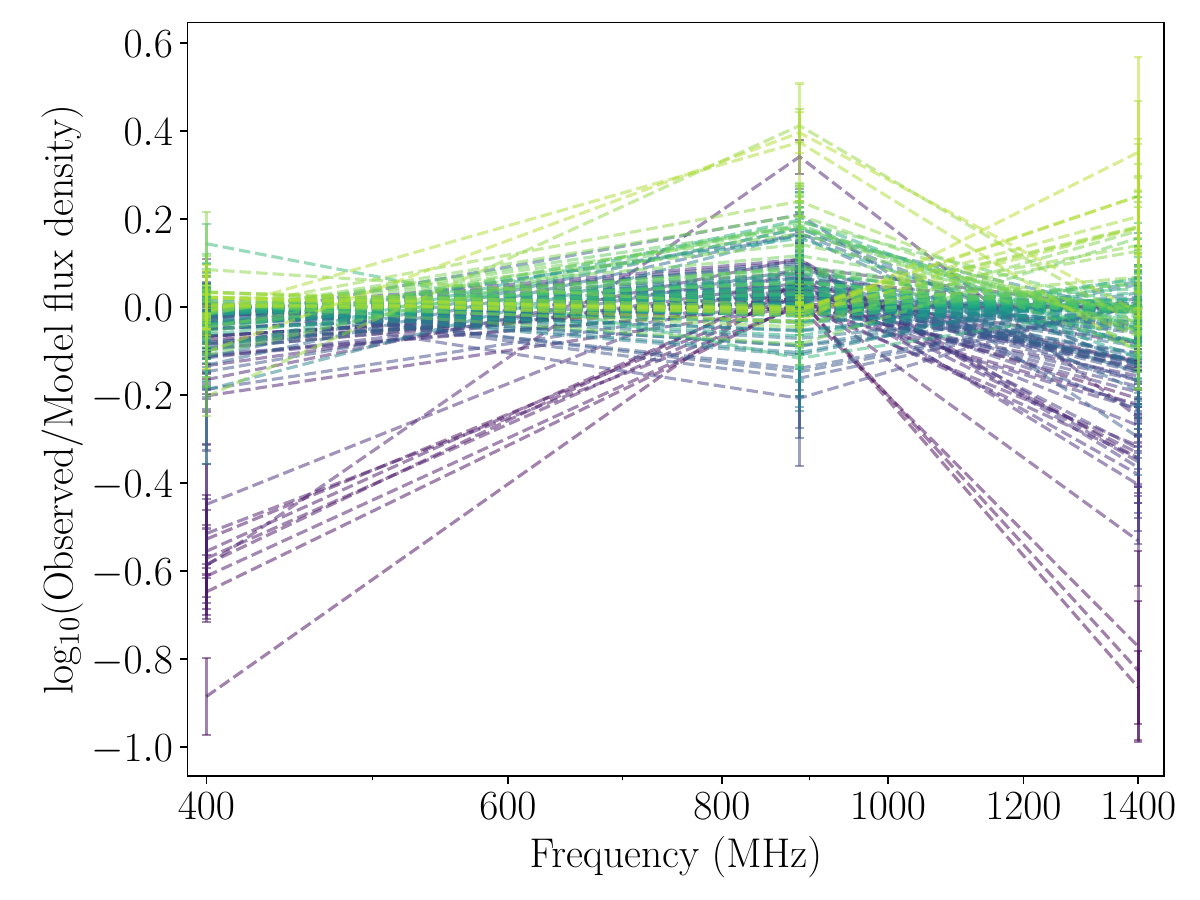}
    \caption{The residual (observed$-$predicted in logarithmic space) or the ratio between the observed flux density and the power-law modeled flux density for the 150 pulsars that exhibit monotonic trends. The individual colored dashed lines show the residuals for each pulsar in our 150-pulsar sample. The color scheme represents the degree of error summed over the three frequencies resulting from a power law fit, with violet and yellow representing maximal over-prediction and under-prediction respectively by a power law spectrum. 
    %Shown in the overlaid solid-colored lines are the representative cases for different spectral variations: i) the red line represents a source spectrum that can be well modeled by a power law, ii) the orange/magenta line shows the case where a power law over/under predicts the flux density at 888\,MHz, iii) the blue/green line shows the case of high-frequency excess/deficit compared to a power law prediction and iv) the purple line shows the case of low-frequency deficit compared to a pure power law.
    It can be clearly seen that for many pulsars, a simple power fit is not adequate where the lower and higher frequency deviations are very commonly seen.
    }
    \label{fig:power_law_trouble}
\end{figure}

Figure~\ref{fig:power_law_trouble} shows the ratio of the observed to the power law modeled flux densities (or the difference between the observed and the power law modeled flux densities in logarithmic space) for our sample of 150 pulsars. If a power law accurately models the observed spectral variation, then this ratio has to be consistent with unity within measurement uncertainties and any variation in addition to this reflects the inability of a single power law to model the source spectrum. We find that in $\sim$ 40\% of the pulsars, the source spectrum cannot be well modeled by a simple power law with low and high-frequency deviations seen commonly in this subset of pulsars.
%We find that the observed source spectra exhibit multiple variations with excess/deficit at 888\,MHz (magenta and orange lines in Figure~\ref{fig:power_law_trouble}) over power-law predictions and also show high-frequency excess/deficits (blue and green lines in Figure~\ref{fig:power_law_trouble}) and low-frequency turn-overs (purple line). 

\begin{figure}
    \plotone{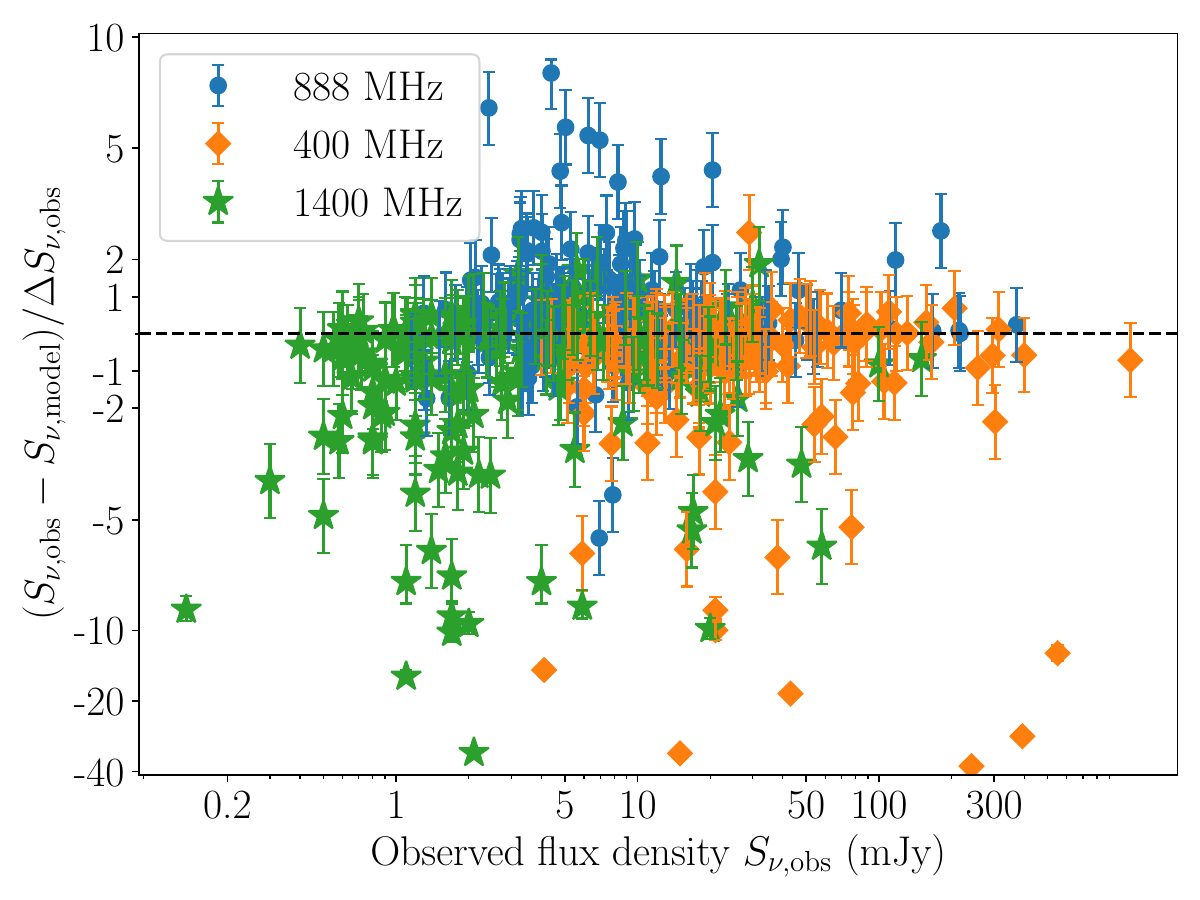}
    \caption{Residuals between the observed flux densities and the power law modeled flux densities at all three different frequencies -- 400, 888, 1400\,MHz, shown in orange, blue, and green error bars respectively. The residuals are normalized by the uncertainty in the flux density measurements at the three different frequencies. 
    %We see that there are many cases where the residuals deviate from the 1-$\sigma$ noise limit showing that a power law is not an adequate fit in these cases.}
    }
    \label{fig:resids}
\end{figure}

The residuals at the three different frequencies are shown in Figure~\ref{fig:resids}. All the residuals are scaled with the flux density uncertainties to see the deviation from white noise. As can be seen, in many cases, the residuals are much larger than the usual 1-$\sigma$ limit (the median value of the residuals is $\sim$ 1.8-$\sigma$, 1.3-$\sigma$, 2-$\sigma$ at 400, 888, 1400\,MHz respectively over/under-predicting the flux at 400 and 1400\,MHz in many cases.  

We note, though, that these measurements are not simultaneous, so temporal variations could appear as spectral variations.  Aside from significant intrinsic variability which is present in some pulsars \citep[e.g.,][]{Kramer2006}, RISS can also cause long-term intensity variations.  However, as shown in \S\ref{sec:scint}, we expect this to be $\sim$6\% -- 18\% for most of the pulsars. There can be a few cases where the fluctuations due to RISS may be comparable to the deviations from a simple power law, but the prevalence  of pulsars in which we see deviations from a simple power law and the large residuals from a power law fit means that RISS alone cannot be responsible. This echoes previous conclusions that a power law is not always a good description of the pulsar spectrum and highlights the low and high-frequency turn-overs that are commonly seen \citep{Maron2000,handbook,MWA2017,Swainston2022}. 

\begin{figure}
    \plotone{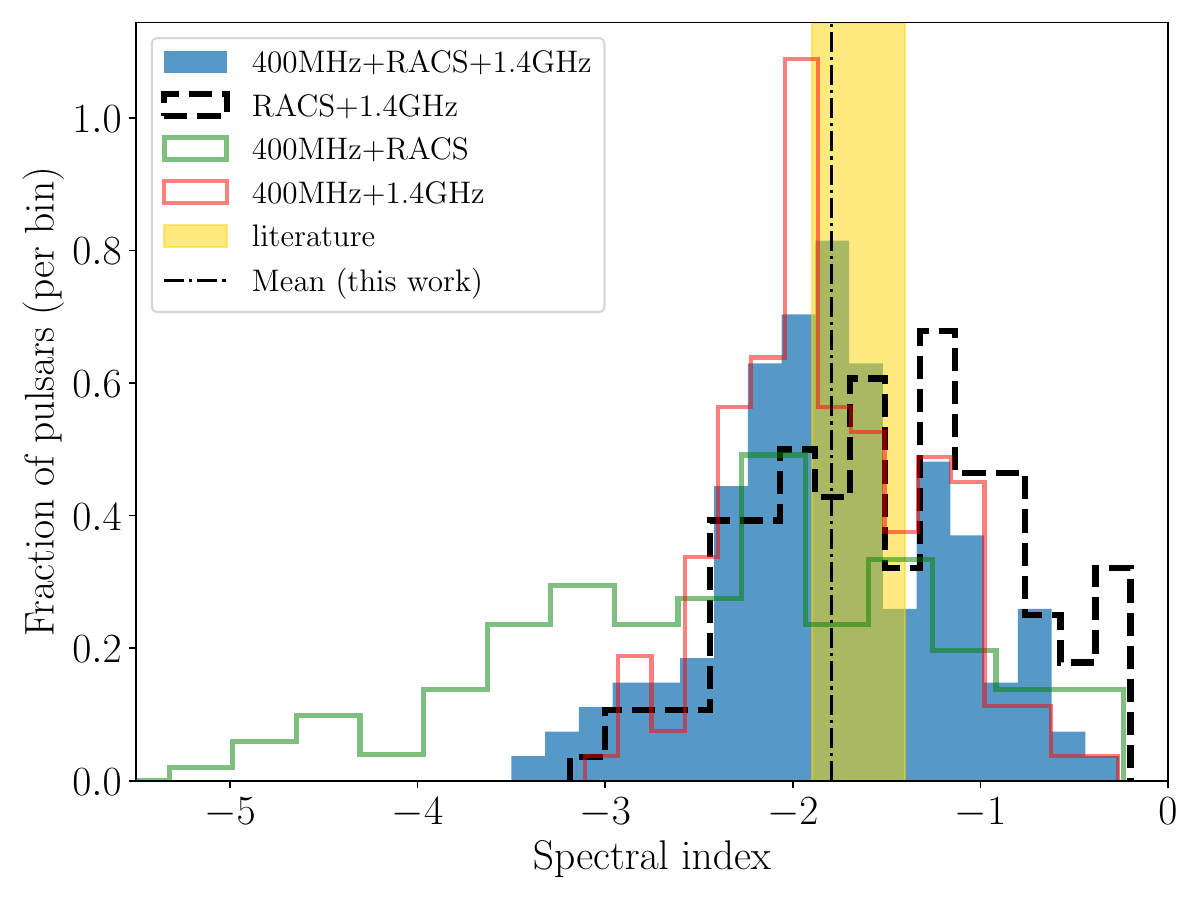}
    \caption{Distribution of spectral indices using the sample of pulsars in RACS that have measured flux densities at both 400 and 1400~MHz. The blue-filled histogram shows the spectral index from a fit using the flux at all three frequencies while the green histogram shows the best-fit index using the two higher frequencies, the dashed black histograms show the same using the two lower frequencies, and the red histogram shows the distribution using archival (400 and 1400\,MHz) data. The golden-filled region shows the range of spectral values reported in the literature (varying from $-$1.9 to $-$1.4) and the black dashed line shows the mean of the distribution using the three frequencies in this work, $-$1.78.}
    \label{fig:spec_fit}
\end{figure}

Figure~\ref{fig:spec_fit} shows the distribution of spectral indices in this sample of 150 pulsars. 
%The blue-filled histogram shows the distribution of spectral indices when the fluxes at all three different frequencies are used, the green histogram shows the fit when the two higher frequencies (888 and 1400\,MHz), the dashed black histogram shows the fit when two lower frequencies are used, and the red histogram shows the distribution when the archival data (400 and 1400\,MHz) are used. 
We find that using two of the three frequencies (the two lower or two higher) for the fit results in different distributions. This can be explained if the source spectrum deviates from the power law in the presence of low/high-frequency deviations.
In the presence of low/high-frequency deviations, using the two lower or two higher frequencies can result in shallower and steeper fits compared to the actual spectrum leading to the deviation between these histograms in figure~\ref{fig:spec_fit}. We find that the mean spectral index is $-$1.78$\pm$0.6, which is towards the steeper end, but still consistent with existing literature \citep[e.g.,][]{ Lorimer1995, Bates2013, jankowski18}.

\begin{figure}
    \plotone{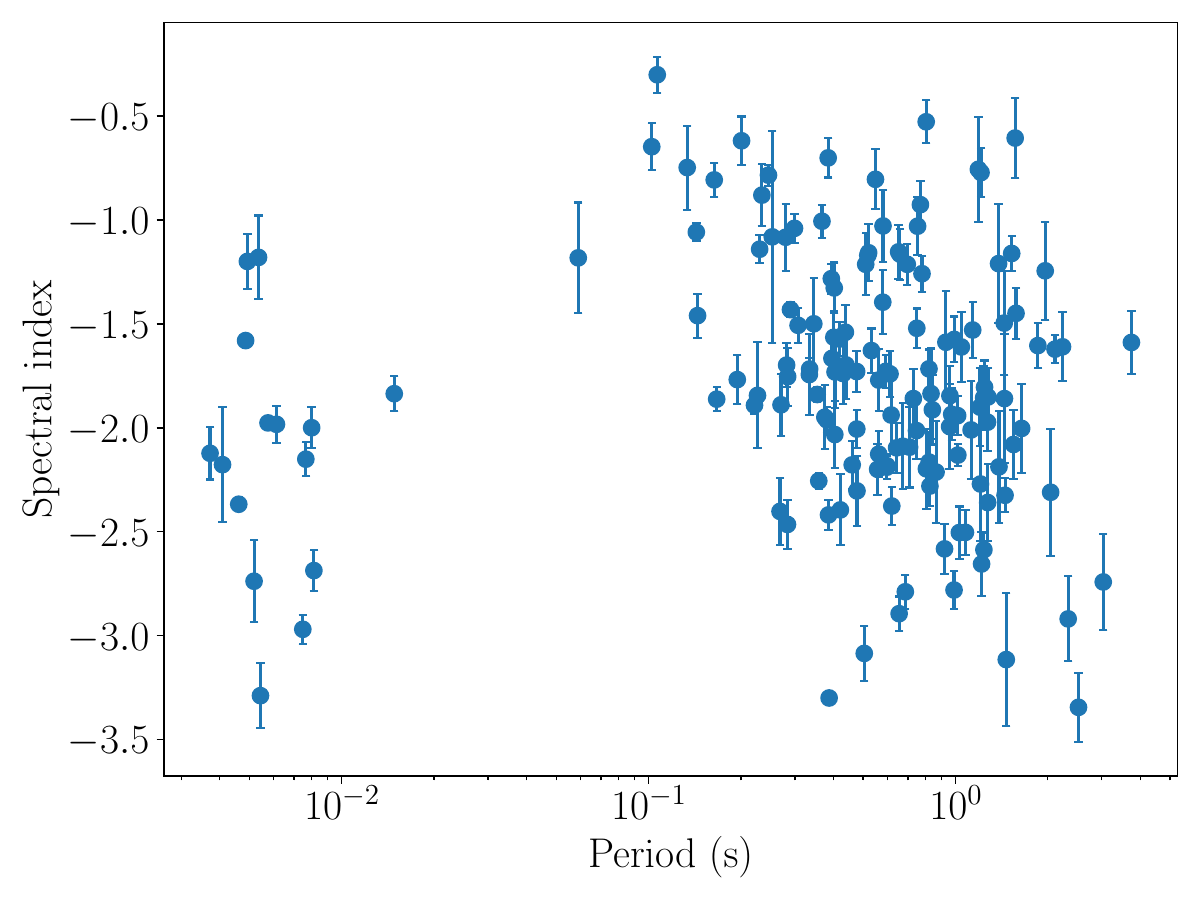}
    \caption{The distribution of spectral indices as a function of the pulsar's period. We see two clear populations representing the normal pulsars and the recycled pulsars. There is no clear evidence for steep spectral indices in recycled pulsars compared to normal pulsars.}
    \label{fig:index_per_corr}
\end{figure}

Figure~\ref{fig:index_per_corr} shows the correlation between the spectral index and the pulsar's period.  We do not see any strong evidence for spectra in recycled pulsars being steeper than the ones in normal pulsars (supported by a p-value of 0.15 from a 2-sample AD test), consistent with past studies \citep{Kramer1998, handbook}.

\subsection{Polarization fraction}

\begin{figure}
    \centering
    \plotone{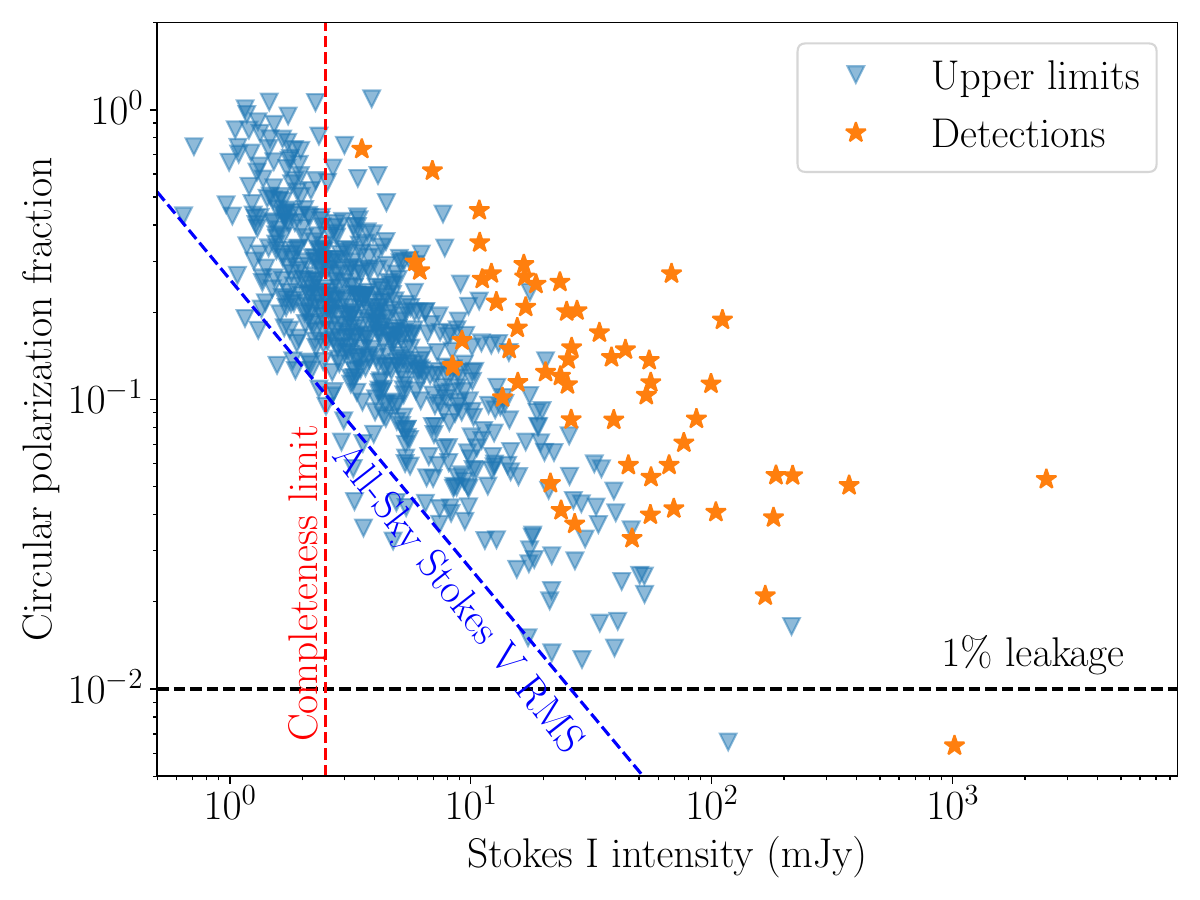}\\
    \plotone{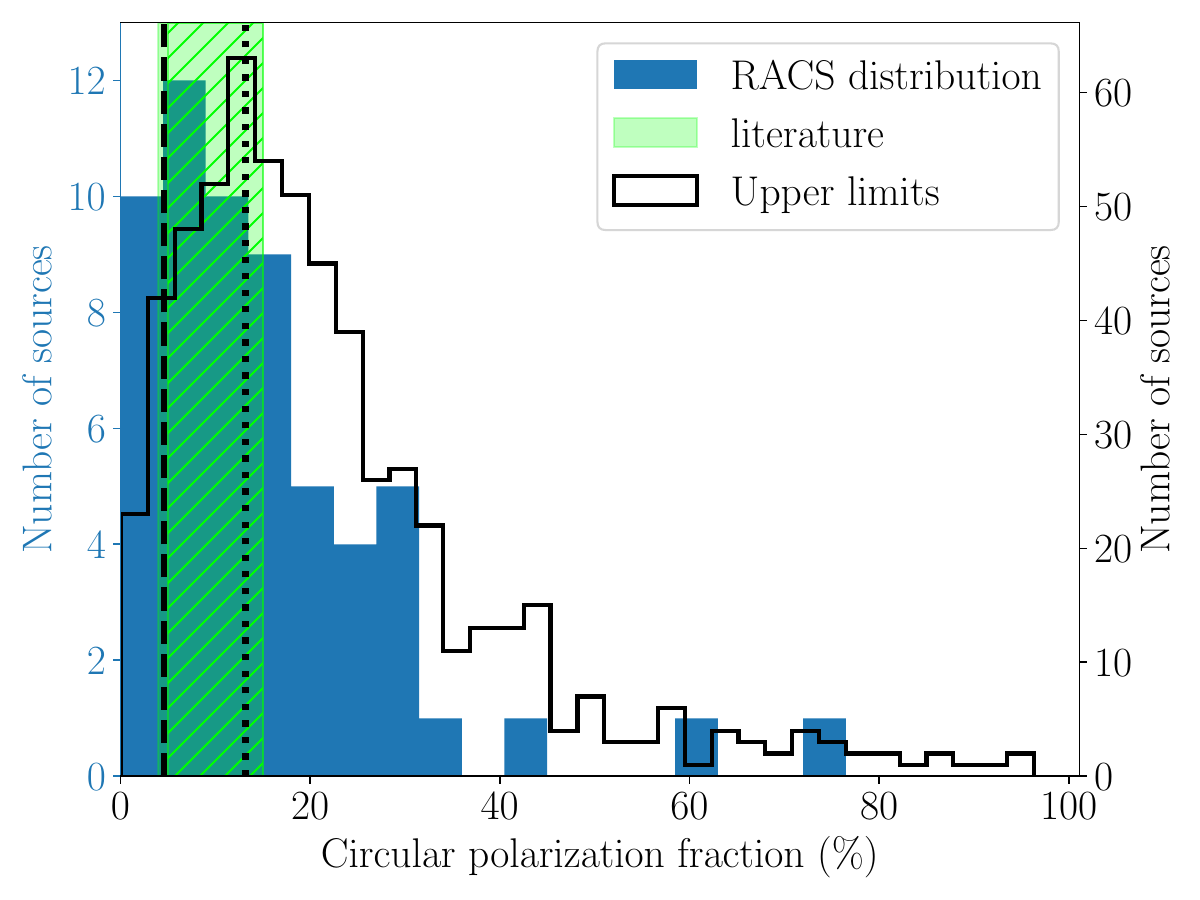}
    % \fig{figures_mod/pol_frac.pdf}{0.5\textwidth}{}\\
    % \vspace{-1cm}\fig{figures_mod/pol_frac_hist.pdf}{0.5\textwidth}{}
    \caption{\textit{Top:} Circular polarization fractions for the pulsars in our sample. Shown in the orange diamonds are the polarization fractions for sources detected in both Stokes I and Stokes V. Shown in the blue triangles are the upper limits in the case of Stokes V non-detections and shown as the black dashed line is the 1\% polarization resulting from leakage into Stokes V \citep{Joshua2021}. Shown in the vertical red dashed line is the Stokes I completeness limit (see Figure~\ref{fig:flux_dist}) and in the blue dashed line is the expected polarization fraction for a Stokes V flux density is twice the all-sky RMS value in the RACS survey. \textit{Bottom:} Distribution of circular polarization fraction for pulsars detected in RACS Stokes V survey. The blue histograms show the distribution for the 61/661 pulsars that are detected in Stokes V images, the black histogram shows the distribution of polarization fraction from non-detections (using upper limits), and the green-filled region shows the polarization values from the literature \citep{Gould1998, Sobey2021, Oswald2023}. The distribution is peaked around 10\% with few pulsars having higher polarization fractions. The black dashed line shows the mean polarization fraction (4.6\%) using the Kaplan-Meier estimate and the black dotted line indicates the $1-\sigma$ spread (8.6\%) from the mean of the distribution.}
    \label{fig:pol}
\end{figure}

We measured the circular polarization fraction in pulsars where we have simultaneous detections of the source in Stokes I and Stokes V (61/661). Polarization fractions range from 0.5\% to $\sim$70\%. For sources that were not detected in Stokes V, we report the upper limits.
The polarization fraction of the matches and upper limits can be seen in Figure~\ref{fig:pol} (top panel). 
Note that polarization fractions below 1\% may not be reliable:
\cite{Joshua2021} showed that circular polarization of $\sim$1\% (twice the median value reported by \citealt{Joshua2021}) can be observed due to the leakage of flux into Stokes V in the RACS data, somewhat dependent on position in the image. 
Other sources with upper limits to the polarization fraction are generally consistent with twice the all-sky Stokes V sensitivity limits, as seen in Figure~\ref{fig:pol}.
%If the sky were uniformly polarized with a Stokes V flux that is twice the RMS value (95\% confidence) of the RACS survey, then any source with a lower polarization would not be detected. 
The black dashed line in Figure~\ref{fig:pol} shows the leakage cutoff and it can be seen that the majority of Stokes V detections are above this leakage level while also being above the Stokes V sensitivity threshold. 
%The dashed blue line shows this limit and the dashed red line shows the survey completeness limit for the total intensity (see \S\ref{sec:completeness} and figure~\ref{fig:flux_dist}). 

For the Stokes V detections, the distribution of polarization fractions is shown in the bottom panel of Figure~\ref{fig:pol}.
We find that most of the pulsar detections in the RACS survey have polarization fractions (median of $\sim$10\%) that are consistent with pulsar observations in the literature \citep[shown in the green stripe in Figure~\ref{fig:pol}][]{Gould1998, Sobey2021, Oswald2023} with a handful of them having higher polarization fractions.
This howver does not take into account the non-detections (upper limits) that dominate the sample ($\sim$ 90\% of the sample). In the presence of a combination of detections and upper limits, we follow \citet{Feigelson1985} to calculate the Kaplan-Meier estimator \citep{kaplanmeier} for the left-censored data (upper limits) and then estimate the mean of the combined data (detections and upper limits). We find that the polarization fractions $<13\%$ are more likely (mean polarization fraction of 4.6\% with a large spread of $\pm$8.4\%) with an extended tail towards the higher values, using the Kaplan-Meier estimator, consistent with the median of the observed distribution.

% To test the hypothesis whether the polarization fraction and the Stokes I intensity are independent (see Figure \ref{fig:pol} \textit{top panel}), we use a non-parametric test, the Kendall rank correlation test \citep{kendalltau}. In the presence of censored data (upper limits), we follow \cite{helsel2005nondetects}, which proposes a generalization to the Kendall test. We find strong evidence (p-value of 0.006) against the null hypothesis of  the absence of any association between the polarization fraction and the Stokes I intensity, suggesting that the two are not independent.

% \begin{figure}
% \includegraphics[width=0.49\textwidth]{figures-created/Polarization_Fraction.pdf}
% \caption{The polarization fraction of pulsars versus their Stokes I flux density. The solid red line is the Stokes V sensitivity of 1.55 mJy divided by the Stokes I flux density, and the green dashed line is 1\% leakage. }
% \label{Fig:PolarizationFractions}
% \end{figure}

\section{Discussions}\label{sec:discussion}
\subsection{Luminosity correlations}\label{sec:corr}
If the underlying source spectrum of a pulsar is known, we can calculate the expected luminosity in a given frequency band. However, this is usually not the case, since we do not know the pulsar's intrinsic emission spectrum. In addition, pulsar emission is beamed and the emission geometry is not well constrained, so following literature \citep[see][for a review]{Stollman87, Bagchi2013}, we define the \textit{pseudo luminosity} as $S_{\nu} {d}^2$, in the units of mJy-kpc$^2$, where $S_{\nu}$ is the flux density at a frequency $\nu$ and $d$ is the distance to the pulsar estimated using the DM and a Galactic electron density model \citep{NE2001, YMW16}. We computed the pseudo luminosity for all the pulsars in our sample using the measured flux densities at 888~MHz to look for any trends that the radio luminosity exhibits with the pulsar's parameters. In general, the radio luminosity function for pulsars is expressed as $L_{\nu}=A\,P^{\gamma_1}\,\dot{P}^{\gamma_2}$ and the indices ($\gamma_1\,, \gamma_2$) are estimated from observations \citep{gunn79, Proszynski84, Stollman87, Bagchi2013}. \cite{gunn79} proposed that the radio pseudo luminosity goes as $B^2\propto P \dot{P}$, \cite{Proszynski84} found that it goes as $P^{-1} \dot{P}^{0.35}$ roughly corresponding to $\dot{E}^{1/3}$, where $\dot{E}$ is the spin-down luminosity, and \cite{Stollman87} proposed a $B^2/P$ dependence, for pulsars with magnetic field $<10^{13}$\,G, pulsars that are dominant in our sample. Hence, we look for any correlations between the pseudo-luminosity and the pulsars' intrinsic parameters (period/period derivative/characteristic age/magnetic field) and the quantities proposed in the literature.

%In addition to these, we defined the radio luminosity fraction as the ratio of radio luminosity observed in the RACS spectral band (defined as the pseudo luminosity $\times$ RACS bandwidth) to the spin-down luminosity.

\begin{figure*}[!htb]
    \plotone{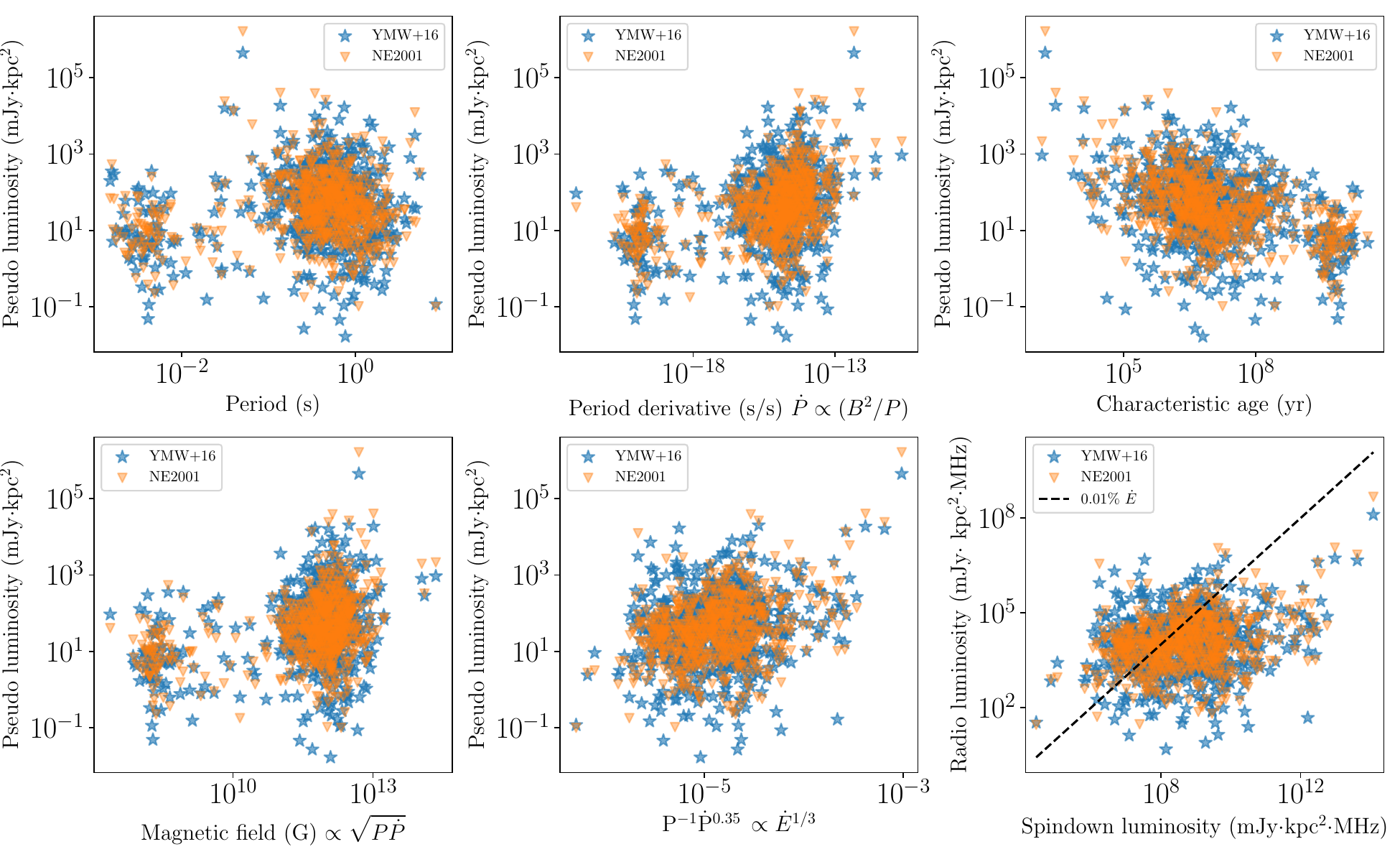}
    \caption{Correlation plots looking for any intrinsic correlation between the pseudo luminosity and the pulsar's spin parameters, characteristic age, the surface dipole magnetic field strength, and other combinations of these parameters of interest (see \S\ref{sec:corr}). Distance to the pulsar is estimated using the electron density models -- the blue scatter plot shows the estimates using \citet{YMW16} and the orange scatter plot shows the same using \citet{NE2001}. A clear distinction between normal pulsars and recycled pulsars can be seen; especially in the spin parameter(s) space. No strong correlations between the pseudo luminosity and any of the pulsar parameters are clearly seen. The bottom-right plot shows the correlation between the radio luminosity (see \S\ref{sec:corr}) and the spin-down luminosity. The black dashed line corresponds to 0.01\% of spin-down luminosity.
    %, except for the radio luminosity fraction (see \S\ref{sec:corr}) with spin-down luminosity.}
    }
    \label{fig:lum_trends}
\end{figure*}

% \begin{figure}[!htb]
%     \centering
%     \includegraphics[scale=0.4]{figures_mod/lum_compare.pdf}
%     \caption{Comparison of the equivalent luminosity vs the spin-down luminosity. Equivalent luminosity is estimated using the RACS flux density at 888~MHz and the DM distance. The blue stars show the DM distance estimate using \cite{YMW16} and the orange triangles show the distance estimate using \cite{NE2001}.}
%     \label{fig:lum_compare}
% \end{figure}

Figure~\ref{fig:lum_trends} shows the correlation plots of pseudo luminosity vs the pulsar's parameters. The blue and the orange scatters show the luminosity estimated using \cite{YMW16} and \cite{NE2001} electron density maps respectively. In all the cases, we found no clear evidence for any strong correlation with the estimated \textit{pseudo luminosity}. To compare this with the spin-down luminosity\footnote{The spin-down luminosity is estimated assuming a moment of inertia $I=10^{45}\,{\rm g\,cm}^2$. We also scale this by 4$\pi$ taking into account the uncertainty in the emission geometry}, we compute radio luminosity (defined as the pseudo luminosity $\times$ RACS bandwidth). We find that this radio luminosity (see the bottom-right panel of figure~\ref{fig:lum_trends}) does not scale accordingly with the spin-down luminosity. The black dashed line shows the expected radio luminosity if it were powered by 0.01\% of the spin-down luminosity implying that varying fractions of spin-down luminosity power the radio emission. We also find that the luminosity ratio (radio to spin-down) decreases with increasing spin-down luminosity and hence increasing fraction of spin-down luminosity powers the radio emission in pulsars as the pulsar ages. 
%The black dotted line shows the equality of spin-down luminosity and the radio luminosity in the RACS spectral band. The major uncertainty in estimating the radio luminosity comes from the DM-based distance to the pulsar. The existence of data points near and beyond this line can be attributed to the underestimation of pulsar distances (??).

% However, when we look at the luminosity ratio, defined as the ratio between the equivalent luminosity to the spin-down luminosity, it seems to be positively correlated with the pulsar period and the characteristic age and negatively correlated with the period derivative as shown in the figure~\ref{fig:lum_rat_trends}. It can also be seen from figure~\ref{fig:lum_rat_trends}, the existence of two classes -- normal and recycled pulsars. We filter normal pulsars visually \fixme{specify} since the initial conditions in the period space for recycled pulsar is unclear and hence can be confused with their evolution. The black dashed lines in figure~\ref{fig:lum_rat_trends} shows these cut-offs.

To quantify the level of this correlation, we use a non-parametric correlation test, the Spearman rank correlation test. Table~\ref{tab:correlations} shows the Spearman correlation coefficients for the pseudo luminosity vs the intrinsic pulsar's parameters and the existing correlations in the literature. Our sample is mainly dominated by normal pulsars, as evident from Figure~\ref{fig:lum_trends}, and hence we restrict our correlation test to normal pulsars (we use the following cuts to distinguish normal from recycled pulsars --- $P=100$\,ms, $\dot{P}=10^{-17}{\rm s\,s}^{-1}$, $\tau_c=300$\,Myr and $B=5\times 10^{10}$\,G). As expected from Figure~\ref{fig:lum_trends}, we do not find any strong evidence for the pseudo luminosity being correlated with any parameter.
%, except for the radio luminosity fraction being negatively correlated to the spin-down luminosity, implying that the radio emission is powered by increasing fractions of spin-down luminosity as the pulsar ages.

\begin{deluxetable}{c|Dr|Dr}
\tablehead{
\colhead{Paramter} & \multicolumn{6}{c}{Correlation coefficient}\\
\tableline
\colhead{DM Model} & \multicolumn{3}{c}{\cite{YMW16}} & \multicolumn{3}{c}{\cite{NE2001}}\\
\colhead{} & \multicolumn{2}{c}{Coefficient} & \colhead{p-value} & \multicolumn{2}{c}{Coefficient} & \colhead{p-value}
}
\decimals
\startdata
$P$ & $-0.15$ & $4\times 10^{-04}$ & $-0.21$ & $5\times 10^{-07}$ \\
$\dot{P}$ & $0.16$ & $1\times 10^{-04}$ & $0.30$ & $2\times 10^{-13}$ \\
$\tau_c$ & $-0.35$ & $9\times 10^{-21}$ & $-0.48$ & $2\times 10^{-38}$ \\
%$B$ & $0.07$ & $9\times 10^{-02}$ & $0.16$ & $8\times 10^{-05}$ \\
$B^2\propto(\rm P \dot{P})$ & $0.07$ & $9\times 10^{-02}$ & $0.17$ & $7\times 10^{-05}$ \\
$P^{-1} \dot{\rm P}^{0.35}\propto \dot{E}^{1/3}$ & $0.17$ & $1\times 10^{-05}$ & $0.28$ & $2\times 10^{-13}$ \\
%$B^2 \rm P^{-1}$ & $0.16$ & $1\times 10^{-04}$ & $0.30$ & $2\times 10^{-13}$ \\
%$\dot{E}_{\rm RACS}$ & $-0.78$ & $<1\times 10^{-99}$ & $-0.81$ & $<1\times 10^{-99}$ \\
\enddata
\caption{Correlation coefficients for the correlation between the pseudo luminosity (see \S\ref{sec:corr}) and the pulsar parameters and existing correlations in the literature. We do not find a strong correlation in any of the above cases.}
\label{tab:correlations}
\end{deluxetable}

% \begin{figure*}[!htb]
%     \centering
%     \includegraphics[scale=0.54]{figures_mod/luminosity_ratio_variations.pdf}
%     \caption{Correlation plots, similar to figure \ref{fig:lum_trends}, but looking for correlation between the observed to spin down luminosity ratio vs the pulsar's intrinsic parameters. Contrary to figure~\ref{fig:lum_trends}, here we see hints of luminosity ratio positively correlated with period and the age and negatively correlated with period derivative. The dashed black line shows the arbitrary cut-off chosen to separate the normal pulsars.}
%     \label{fig:lum_rat_trends}
% \end{figure*}

\subsection{Comparison with other surveys}
We used the flux measurements from contemporary all-sky radio imaging surveys like the TIFR GMRT Sky Survey \citep[TGSS][at 150\,MHz]{tgss}, Murchison Widefield Array \citep[MWA][roughly at 200\,MHz]{MWA2017} and Very Large Array Sky Survey \citep[VLASS][at 3\,GHz]{vlass} to validate and compare our spectral fits (see \S\ref{sec:spec}). We selected the pulsars where the flux density measurements are available at least at five out of the six different frequencies -- 150, 200, 400, 888, 1400, 3000\,MHz. Using the power law spectra that we computed with the RACS and PSRCAT (\S\ref{sec:spec}), we estimate the predicted flux density at these frequencies and compare it with the corresponding measured flux densities.  

\begin{figure}
    \centering
    \plotone{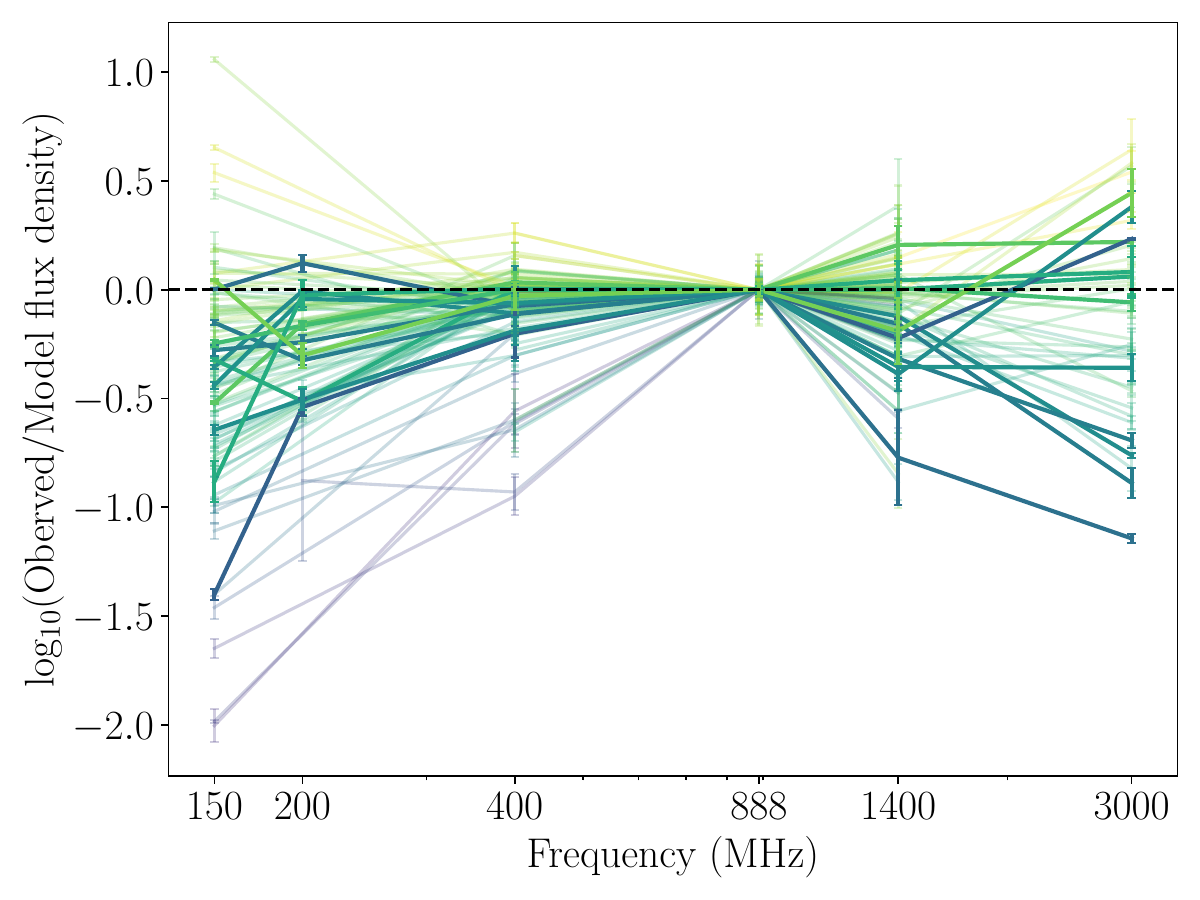}
    \caption{Comparison of the predicted flux densities at 150\, 200, and 3000\,MHz using the spectrum estimated with RACS and PSRCAT data, with the measured flux densities from other imaging surveys. Shown are residuals (in logarithmic space) or the ratio  between the observed fluxes and the modeled fluxes (in linear space); hence any residual structure corresponds to the deviation from a single power law. We observe that in many of the pulsars, a single power law estimated from the ATNF and RACS data (see figure~\ref{fig:power_law_trouble} and \S\ref{sec:spec}) greatly overestimates the flux at these frequencies (while underestimating in a few) hinting the evidence for low and high-frequency deviations from a power law model. The color scheme represents the extent of deviation from a power law (violet/yellow corresponds to maximum deficit/excess). These individual pulsar residuals are adjusted to 0 at 888\,MHz.}
    \label{fig:allresids}
\end{figure}

Figure~\ref{fig:allresids} shows the comparison of the flux densities. Residuals (in logarithmic space) or the ratio  between the observed fluxes and the modeled fluxes (in linear space) (similar to Figure~\ref{fig:power_law_trouble}) are estimated using the power law fit (see S\ref{sec:spec}). If the source spectrum can be well modeled by a single power law, then the residuals are expected to be consistent with zero (the black dashed line) within error limits. However, any additional variation can be interpreted as a single power law being an inadequate description of the source spectrum. We see the evidence for a single-power law greatly overestimating the flux density at the lower and higher frequencies in most of the pulsars (whilst underestimating in a few). 

In the sample of 35 pulsars that have flux density measurements at all five frequencies (150, 400, 888, 1400, 3000\,MHz), we tried to fit for a single power law, this time including all these flux density measurements. 

\begin{figure}
    \centering
    \includegraphics[scale=0.45]{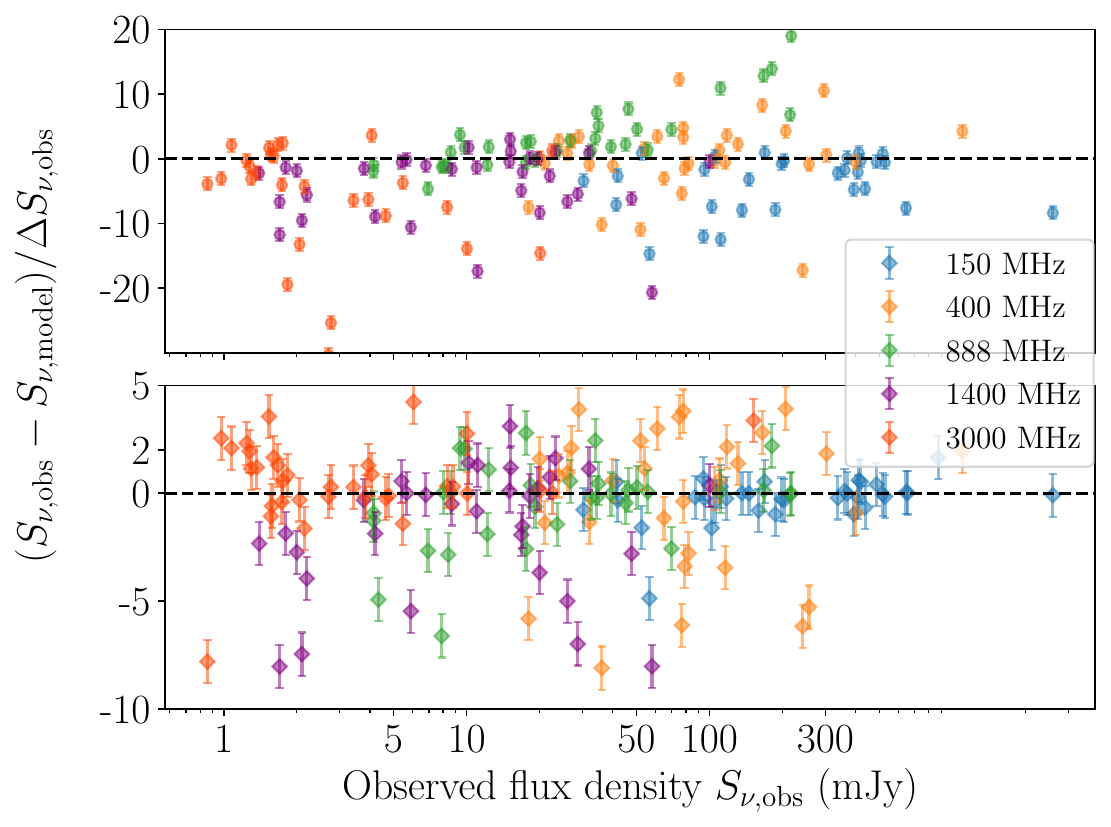}
    \caption{Residuals for the sample of 35 pulsars that have flux measurements at five frequencies (150, 400, 888, 1400, and 3000\,MHz). The top panel shows the residuals when the data are fit using a pure power law and the bottom panel shows the residuals when the data are fit using a quadratic power law. In both cases, the residuals are normalized using the measurement uncertainties.}
    \label{fig:four_freq_fit}
\end{figure}

\begin{figure*}
    \centering
    \includegraphics[scale=0.45]{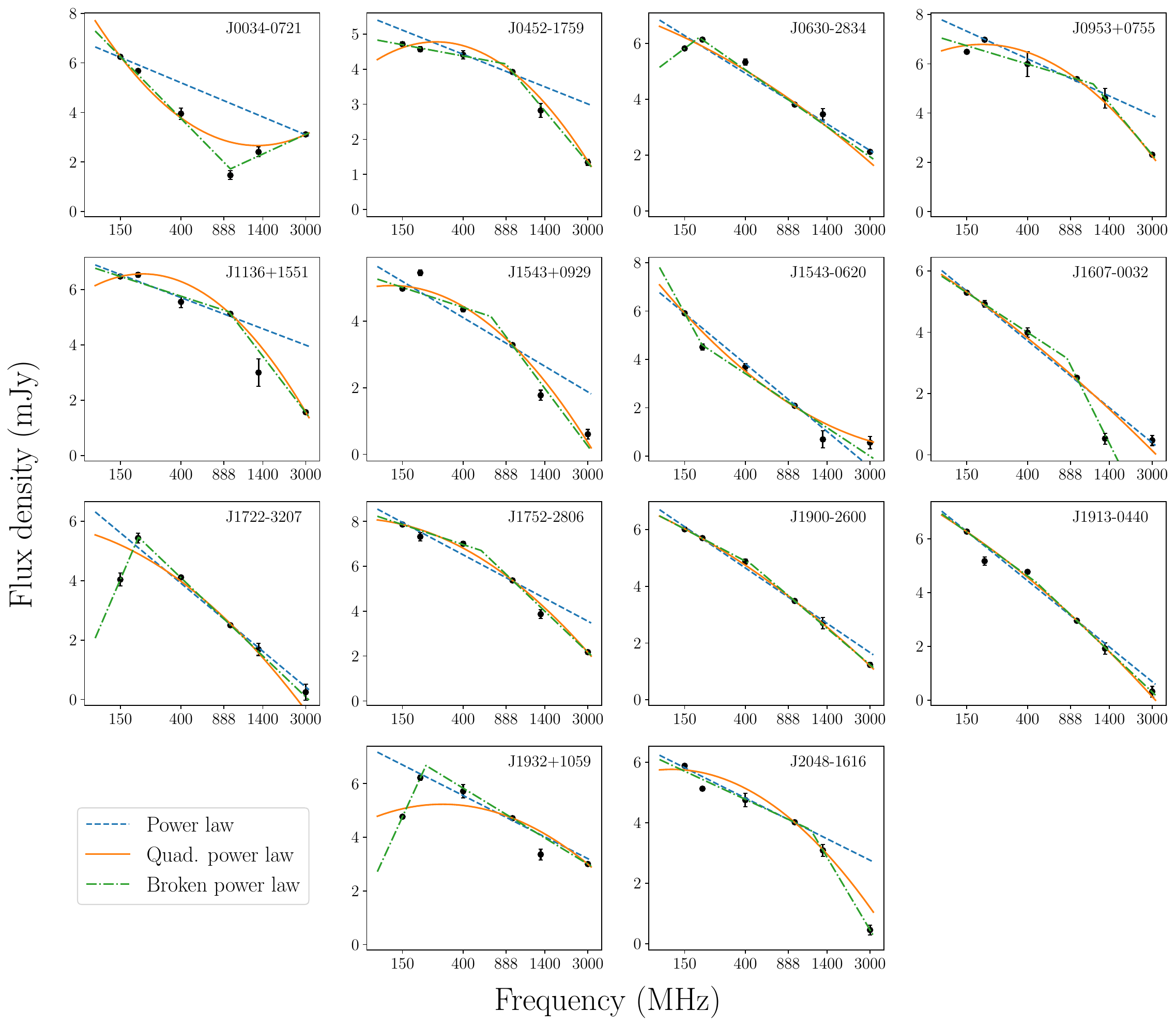}
    \caption{Spectra of the 14 pulsars that have flux density measurements available at 150, 200, 400, 888, 1400, 3000\,MHz. Spectra were modeled using a simple power law, a quadratic power law, and a broken power law. In many cases (e.g., J0452$-$1759) we see that the spectrum prefers a quadratic/broken power law over a simple power law indicating that the deviation from a simple power law is quite common.}
    \label{fig:all_spectra_3_models}
\end{figure*}

Figure~\ref{fig:four_freq_fit} (\textit{top panel}) shows the residuals when the data were fit using a single power law. We find that a single power law does not adequately fit the data, expected from combined RACS and PSRCAT fits (see \S\ref{sec:spec}), with the median residuals (scaled by the measurement uncertainty) $\sim 3\sigma,  3\sigma,  2\sigma, 4\sigma$ at 150, 400, 1400, 3000\,GHz and a median reduced-$\chi^2$ of 7.5 (4 DOF). In this case, we find that the mean spectral index is softer, $-1.53\pm 0.58$, than the estimate derived using the three higher frequencies. We then tried to fit the data using a quadratic power law, 
\begin{equation}
    \ln(S_{\nu}) = a \ln(\nu)^2+b \ln(\nu)+c \nonumber   
\end{equation}
The bottom panel of Figure~\ref{fig:four_freq_fit} shows the residuals in this case and shows that the variation is better modeled by a quadratic power law (uncertainty normalized median residuals are $\sim 0.3\sigma, 2\sigma, 0.6\sigma, 0.8\sigma$, and a median reduced-$\chi^2$ of 1.7 (3 DOF)) rather than a pure power law. In many cases, the spectrum seems to exhibit low-frequency turn-overs \citep{handbook}, and hence a quadratic variation in logarithmic space is able to better capture this trend. We also tried to fit the spectrum using a broken power law and found that it performs comparably to the quadratic power law. Figure~\ref{fig:all_spectra_3_models} shows the spectra of the 14 pulsars that have flux density measurements in TGSS, MWA and VLASS in addition to the RACS and ATNF measurements. We see that the deviation from a simple power law can be quite common with a quadratic power law/broken power law providing a much better fit to the data.

However, we do caution that although a quadratic/broken power law provides a better fit than a single power law, there are still cases where it is still inadequate to model the spectrum; for example, when the spectrum exhibits both low-frequency turnover and high-frequency turn-up, a cubic variation might be needed. In summary, we find that the spectrum in a modest set of pulsars (as large as 40\%) does not seem to exhibit a linear variation (in logarithmic frequency-flux density space) with higher-order non-linear corrections providing better fits, and hence the use of a simple power-law spectral fits in pulsars must be treated with caution.

%The top panel shows the direct comparison of the two, where the black dashed line represents the one-to-one correspondence and the bottom panel shows the residual flux as a function of anticipated flux. A rough trend of the following can be seen: If we use our best-fit model for the source spectrum using the RACS data, then we see that we over-estimate the flux in the pulsars that are supposed to be brighter at 150~MHz and under-estimate the flux in pulsars that are supposed to be less luminous. This can be explained using the low-frequency and high-frequency turnovers observed in pulsar spectra \citep{handbook}. In the presence of a low-frequency turn-down, the power law spectrum overestimates the flux at lower frequencies. In the presence of a high-frequency turn-up, since our estimation of the source spectrum uses the flux density at 1.4~GHz, the source spectrum ends up shallower due to the turn-up which This underestimates the flux at lower frequencies. \textit{But the clear separation of these two populations is not clear, for it means that these frequency turn-overs appear in two different populations??}

\section{Conclusions}\label{sec:conclusion}
We present cross-matches for the known pulsar population against  the first release of the  ASKAP RACS survey data. We find 600 Stokes I sources and 61 sources that have both Stokes I and Stokes V matches to  known pulsars: we expect as many as 0.5\% of these to represent false matches with 95\% confidence. We also present the spectral characterization of these sources finding that a single power law can be inadequate in many cases. Combining this with more low and high-frequency data (TGSS, MWA, VLASS), we find that a quadratic/broken power law represents a better fit to the spectral shape than a pure power law, revealing that the variation of flux density with frequency in logarithmic space can be non-linear and high/low-frequency deviations can be very common. Data presented here can be added to repositories like \citet{Swainston2022} and used in more advanced spectral modeling.  We present the polarization information of these sources finding that the estimated fraction is consistent with the ones in the literature. We looked at the variation of \textit{pseudo luminosity} and its correlation with any intrinsic pulsar parameters and found no significant evidence for a strong correlation, also revealing that varying fractions of spin-down luminosity powers the radio luminosity and this fraction increases as the pulsar ages. The addition of reliable flux density measurements through current/future imaging surveys can help in the accurate modeling of the underlying source spectrum of the pulsars. 

% \begin{acknowledgments}
AA, AE, MJ, and DK are supported by National Science Foundation (NSF) Physics Frontiers Center award numbers 1430284 and 2020265. AA and DK  are further supported by NSF grant AST-1816492. RS is supported by NSF grant AST-1816904.  Parts of this research were conducted by the Australian Research Council Centre of Excellence for Gravitational Wave Discovery (OzGrav), project number CE170100004. This scientific work uses data obtained from Inyarrimanha Ilgari Bundara / the CSIRO Murchison Radio-astronomy Observatory. We acknowledge the Wajarri Yamaji People as the Traditional Owners and native title holders of the Observatory site. CSIRO’s ASKAP radio telescope is part of the Australia Telescope National Facility (\href{https://ror.org/05qajvd42}{ATNF}). Operation of ASKAP is funded by the Australian Government with support from the National Collaborative Research Infrastructure Strategy. ASKAP uses the resources of the Pawsey Supercomputing Research Centre. Establishment of ASKAP, Inyarrimanha Ilgari Bundara, the CSIRO Murchison Radio-astronomy Observatory and the Pawsey Supercomputing Research Centre are initiatives of the Australian Government, with support from the Government of Western Australia and the Science and Industry Endowment Fund. The Parkes radio telescope is part of the Australia Telescope National Facility (https://ror.org/05qajvd42) which is funded by the Australian Government for operation as a National Facility managed by CSIRO.
% \end{acknowledgments}

\facilities{ASKAP}
\software{\texttt{NumPy} \citep{numpy}, \texttt{Matplotlib} \citep{matplotlib}, \texttt{AstroPy} \citep{astropy1, astropy2}}

\newpage

\appendix
\section{Image cutouts}\label{sec:app}

\begin{figure*}[!htb]
    \gridline{
    \fig{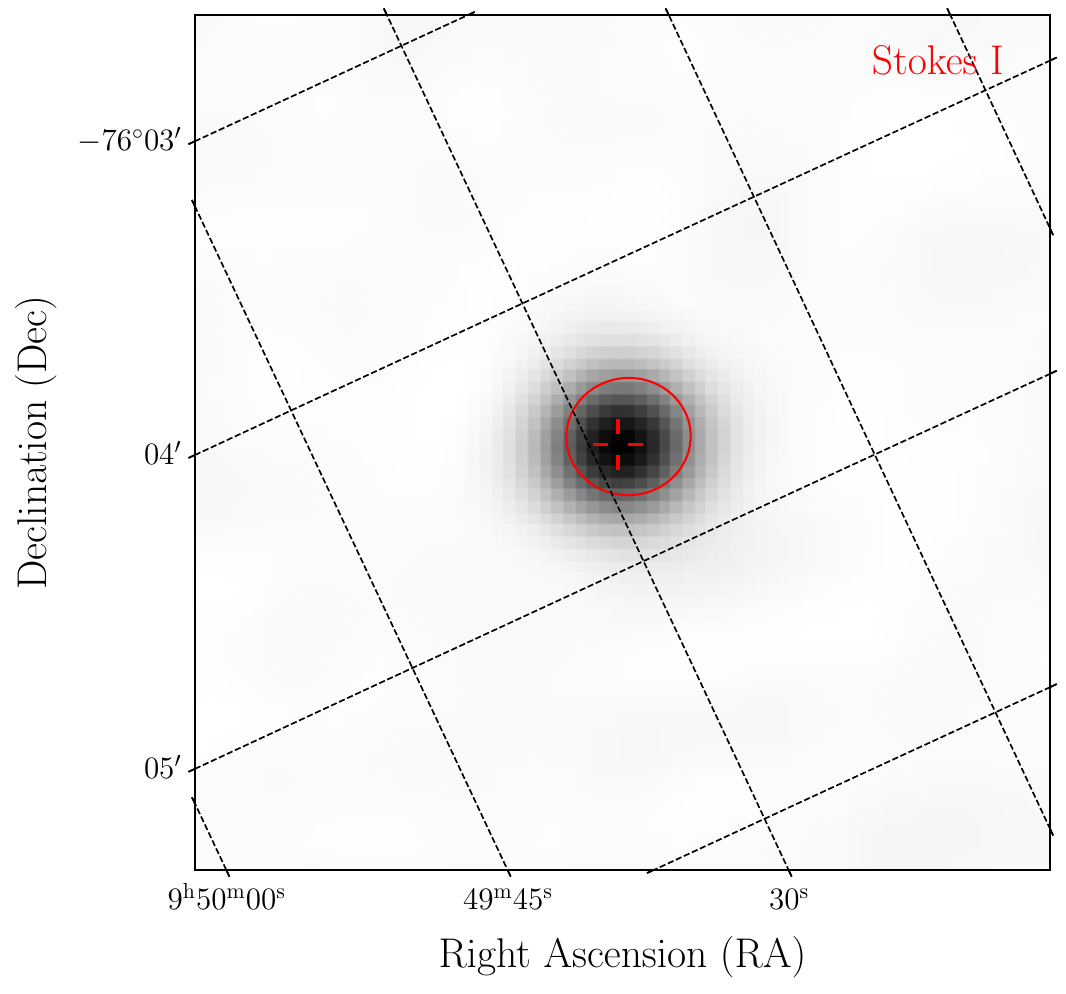}{0.325\textwidth}{(a) PSR J0711$-$6830 (Stokes I)}
    \fig{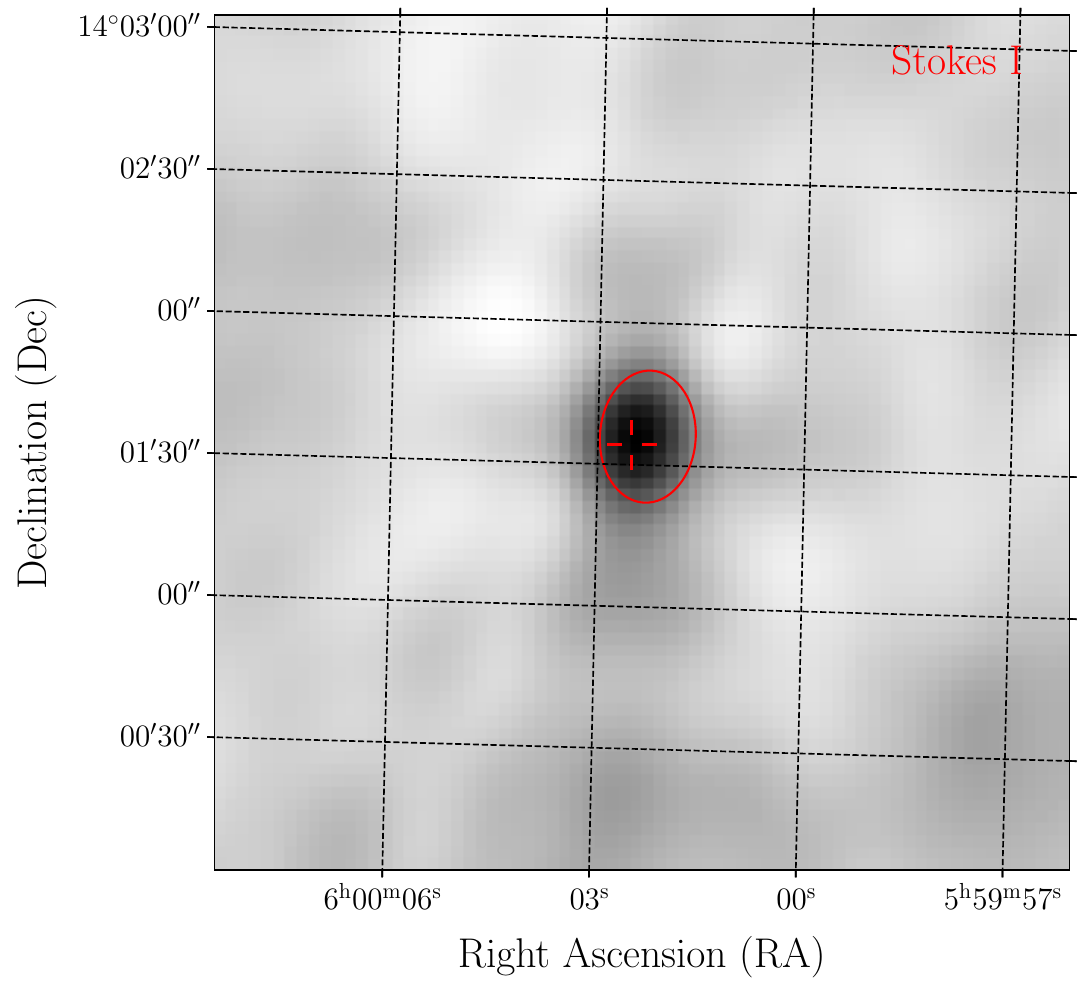}{0.325\textwidth}{(b) PSR J0528+2200 (Stokes I)}
    \fig{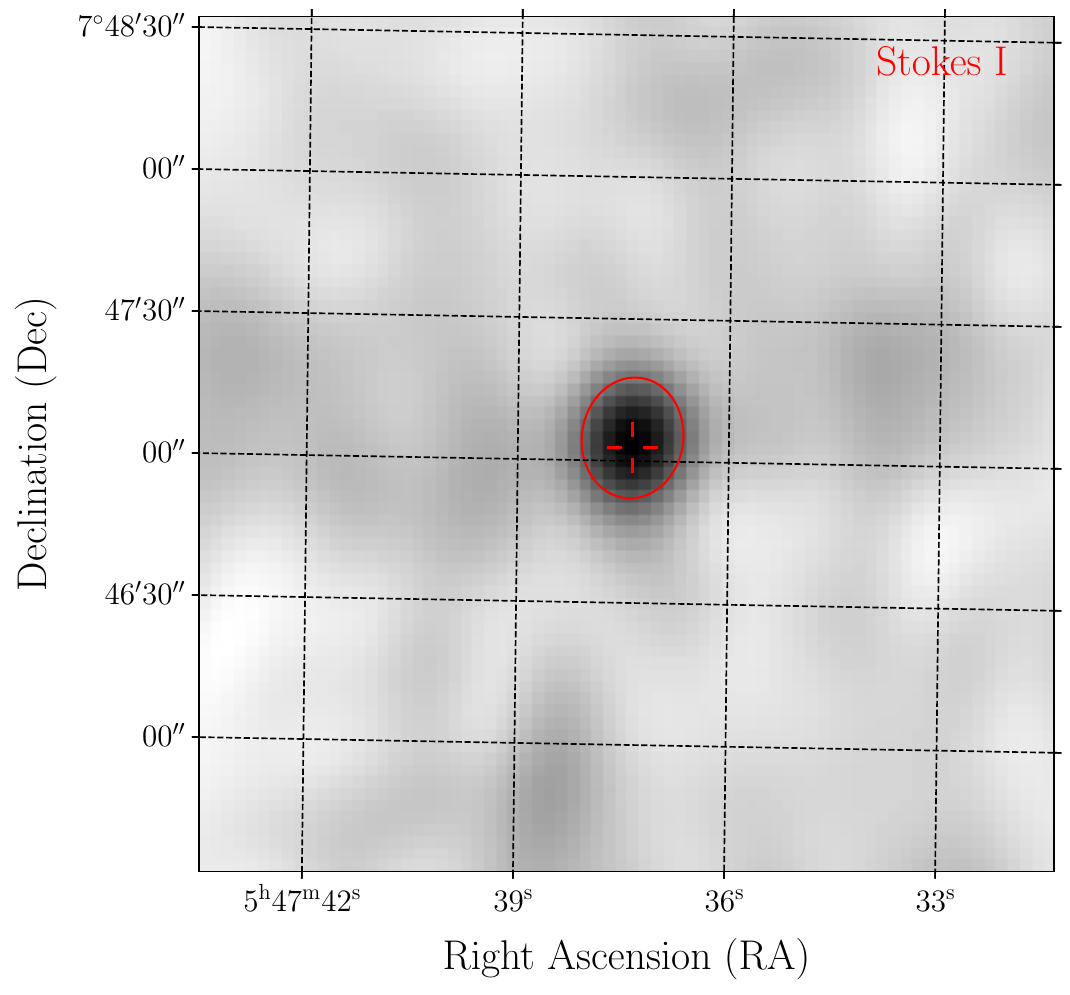}{0.325\textwidth}{(c) PSR J0509+0856 (Stokes I)}
    }
    \gridline{
    \fig{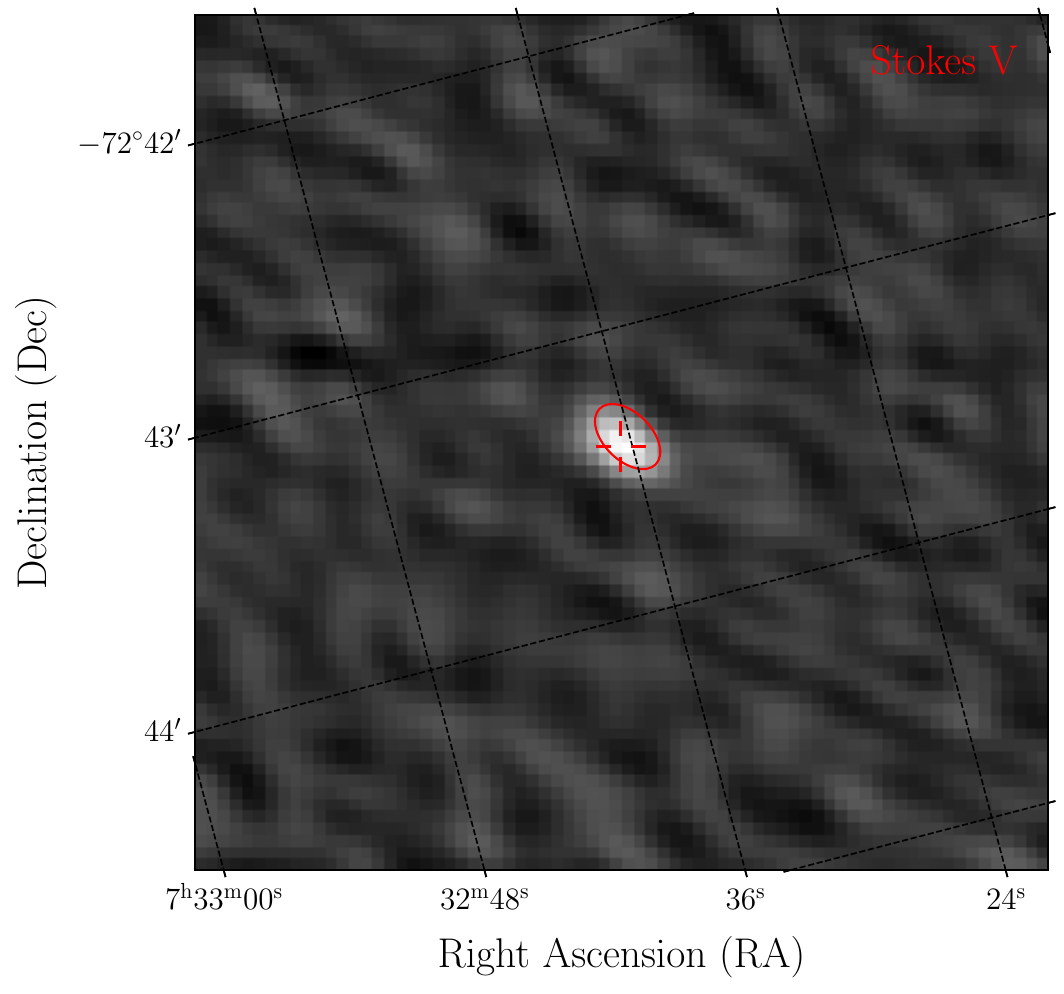}{0.325\textwidth}{(d) PSR J0711$-$6830 (Stokes V)}
    \fig{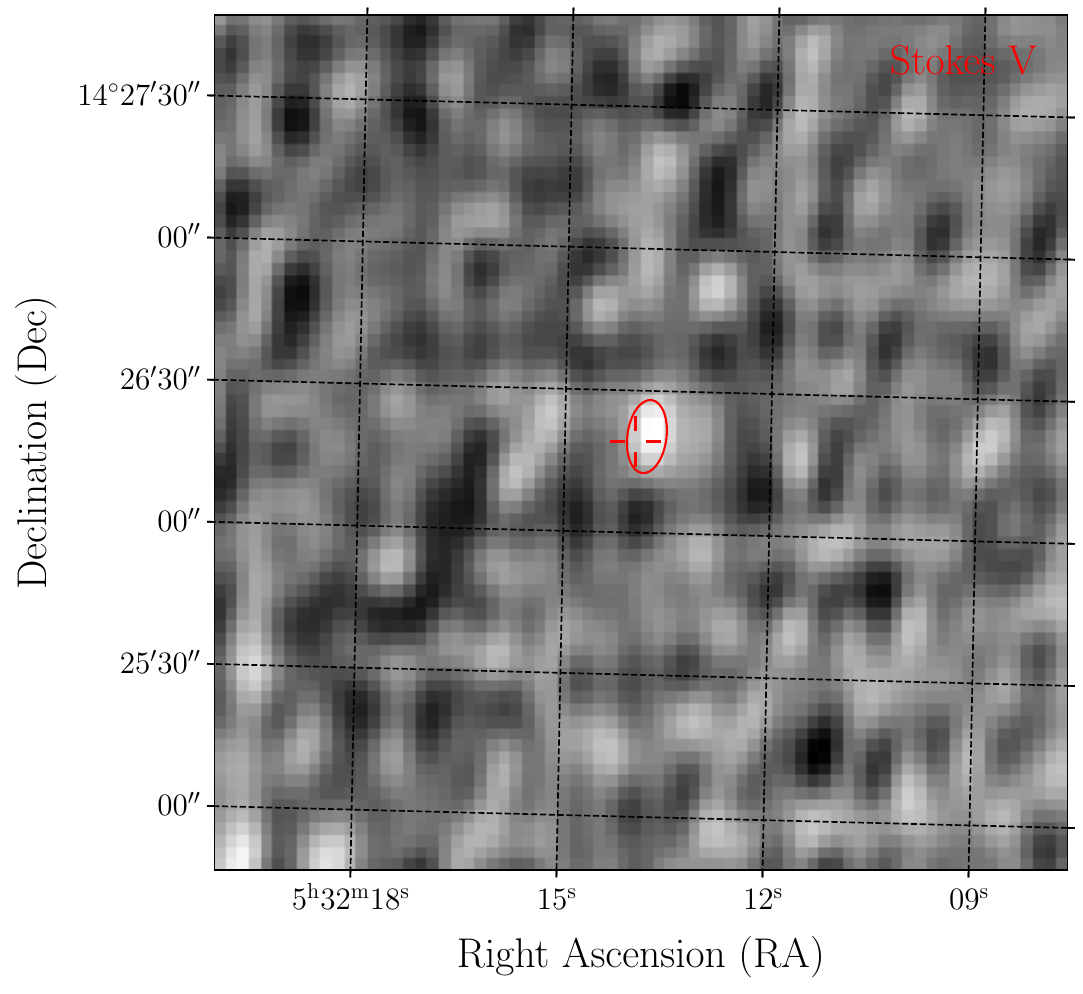}{0.325\textwidth}{(e) PSR J0528+2200 (Stokes V)}
    \fig{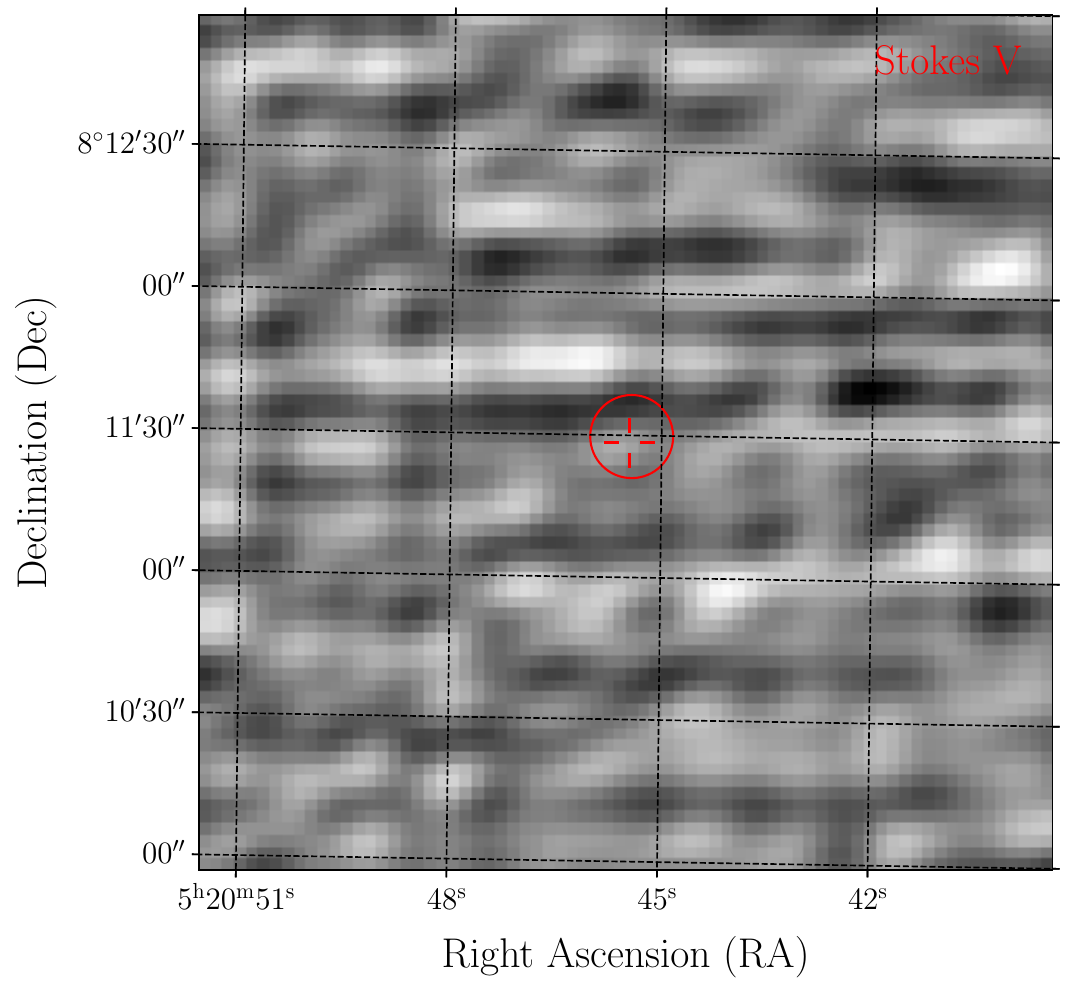}{0.325\textwidth}{(f) PSR J0509+0856 (Stokes V)}
    }
    % \gridline{
    % \fig{figures_mod/racs_images/J0528+2200_racsi.pdf}{0.325\textwidth}{(c) PSR J0528+2200 (Stokes I)}
    % \fig{figures_mod/racs_images/J0528+2200_racsv.pdf}{0.325\textwidth}{(d) PSR J0528+2200 (Stokes V)}
    % }
    % \gridline{
    % \fig{figures_mod/racs_images/J1043-6116_racsi.pdf}{0.4\textwidth}{(e) PSR J1043$-$6116 (Stokes I)}
    % \fig{figures_mod/racs_images/J1043-6116_racsv.pdf}{0.4\textwidth}{(f) PSR J1043$-$6116 (Stokes I)}
    % }
    % \gridline{
    % \fig{figures_mod/racs_images/J0509+0856_racsi.pdf}{0.325\textwidth}{(e) PSR J0509+0856 (Stokes I)}
    % \fig{figures_mod/racs_images/J0509+0856_racsv.pdf}{0.325\textwidth}{(f) PSR J0509+0856 (Stokes I)}
    % }
    \caption{3\arcmin\, image cutouts of Stokes I and Stokes V intensity maps depicting three different cases --- combined Stokes I \& V detection (PSR J0711$-$6830), Stokes I detection with a marginal Stokes V detection (PSR J0528+2200), and a purely Stokes I detection (PSR J0509+0856). Shown in the red ellipse is the detected RACS source and in the cross hairs is the pulsar's position as reported in the ATNF catalog.}
\end{figure*}

\newpage
\clearpage
\section{Flux measurements}
\startlongtable
\begin{deluxetable*}{ccc|ccccccc|l}
\tablehead{\colhead{Pulsar} & \colhead{RA} & \colhead{DEC} & \multicolumn{7}{c}{Flux density\tablenotemark{a}} & \colhead{Index}\\
\tableline
\colhead{} & \colhead{} & \colhead{} & \multicolumn{2}{c}{RACS} & \multicolumn{2}{c}{ATNF} & \multicolumn{3}{c}{Other imaging surveys} & \colhead{}\\
\cline{4-5}\cline{6-7}\cline{8-10}
\colhead{} & \colhead{} & \colhead{} & \multicolumn{2}{c}{888\,MHz} & \colhead{400\,MHz} & \colhead{1.4\,GHz} & \colhead{150\,MHz} & \colhead{200\,MHz} & \colhead{3\,GHz} & \colhead{}\\
\cline{4-10}
% \colhead{} & \colhead{} & \colhead{}  & \colhead{$S_{\nu}$} & \colhead{$\Delta S_{\nu}_{\rm scint}$} & \colhead{$S_{\nu}$} & \colhead{$S_{\nu}$} & \colhead{$S_{\nu}$} & \colhead{$S_{\nu}$} & \colhead{$S_{\nu}$} & \colhead{}\\
\colhead{} & \colhead{} & \colhead{}  & \colhead{(mJy)} & \colhead{(mJy)\tablenotemark{b}} & \colhead{(mJy)} & \colhead{(mJy)} & \colhead{(mJy)} & \colhead{(mJy)} & \colhead{(mJy)} & \colhead{}}
\tablecaption{Astrometric and spectral measurements for pulsars in our sample whose spectrum can be described by a simple power law.}
\startdata
J0030+0451 & 00$^{\rm h}$\;30$^{\rm m}$\;27$\fs$2 & +04$\arcdeg$\;51$\arcmin$\;41$\arcsec$ & \phm{00}2.2(\phm{0}8) & 1.74 & \phm{00}7.9(\phm{00}2) & \phm{00}1.1(\phm{0}3) & \phm{0}45(\phm{0}4) & \nodata & \nodata & $-$1.58(3) \\
J0152$-$1637 & 01$^{\rm h}$\;52$^{\rm m}$\;10$\fs$7 & $-$16$\arcdeg$\;37$\arcmin$\;53$\arcsec$ & \phm{00}4.1(\phm{0}4) & 1.06 & \phm{0}20\phm{.0}(\phm{00}4) & \phm{00}2.1(\phm{0}4) & \phm{0}88(\phm{0}7) & \nodata & \nodata & $-$1.8\phm{0}(2) \\
J0525+1115 & 05$^{\rm h}$\;25$^{\rm m}$\;56$\fs$3 & +11$\arcdeg$\;15$\arcmin$\;20$\arcsec$ & \phm{00}3.7(\phm{}96) & 0.08 & \phm{0}19.5(\phm{00}9) & \phm{00}1.9(\phm{0}2) & \phm{0}32(\phm{0}6) & \nodata & \nodata & $-$1.84(4) \\
J0528+2200 & 05$^{\rm h}$\;28$^{\rm m}$\;52$\fs$2 & +22$\arcdeg$\;00$\arcmin$\;04$\arcsec$ & \phm{0}14\phm{.0}(\phm{0}2) & 0.46 & \phm{0}57\phm{.0}(\phm{00}5) & \phm{00}9\phm{.0}(\phm{0}2) & \nodata & \nodata & \phm{00}1.5(\phm{}4) & $-$1.6\phm{0}(2) \\
J0543+2329 & 05$^{\rm h}$\;43$^{\rm m}$\;09$\fs$7 & +23$\arcdeg$\;29$\arcmin$\;07$\arcsec$ & \phm{0}15.7(\phm{0}7) & 0.22 & \phm{0}29\phm{.0}(\phm{00}1) & \phm{0}10.7(\phm{0}7) & \nodata & \nodata & \phm{00}4.7(\phm{}2) & $-$0.79(5) \\
J0601$-$0527 & 06$^{\rm h}$\;01$^{\rm m}$\;58$\fs$9 & $-$05$\arcdeg$\;27$\arcmin$\;49$\arcsec$ & \phm{00}6.1(\phm{0}4) & 0.05 & \phm{0}22.7(\phm{00}9) & \phm{00}2.6(\phm{0}5) & \phm{0}32(\phm{0}6) & \nodata & \nodata & $-$1.67(8) \\
J0614+2229 & 06$^{\rm h}$\;14$^{\rm m}$\;16$\fs$9 & +22$\arcdeg$\;29$\arcmin$\;58$\arcsec$ & \phm{00}8.4(\phm{0}9) & 0.12 & \phm{0}29\phm{.0}(\phm{00}1) & \phm{00}3.3(\phm{0}2) & \nodata & \nodata & \nodata & $-$1.72(5) \\
J0629+2415 & 06$^{\rm h}$\;29$^{\rm m}$\;05$\fs$6 & +24$\arcdeg$\;15$\arcmin$\;41$\arcsec$ & \phm{00}8.8(\phm{0}8) & 0.12 & \phm{0}31\phm{.0}(\phm{00}2) & \phm{00}3.2(\phm{0}4) & \nodata & \nodata & \nodata & $-$1.7\phm{0}(1) \\
J0659+1414 & 06$^{\rm h}$\;59$^{\rm m}$\;48$\fs$3 & +14$\arcdeg$\;14$\arcmin$\;23$\arcsec$ & \phm{00}3.4(\phm{0}5) & 0.94 & \phm{00}6.5(\phm{00}6) & \phm{00}2.7(\phm{0}2) & \nodata & \nodata & \nodata & $-$0.70(9) \\
J0729$-$1836 & 07$^{\rm h}$\;29$^{\rm m}$\;32$\fs$4 & $-$18$\arcdeg$\;36$\arcmin$\;42$\arcsec$ & \phm{00}4.8(\phm{0}7) & 0.27 & \phm{0}11.2(\phm{00}7) & \phm{00}1.9(\phm{0}5) & \nodata & \nodata & \nodata & $-$1.2\phm{0}(1) \\
J0820$-$1350 & 08$^{\rm h}$\;20$^{\rm m}$\;26$\fs$3 & $-$13$\arcdeg$\;50$\arcmin$\;56$\arcsec$ & \phm{0}12.8(\phm{0}4) & 0.56 & \phm{}102\phm{.0}(\phm{00}6) & \phm{00}6\phm{.0}(\phm{0}2) & \phm{}207(\phm{0}8) & \phm{}160(\phm{0}7) & \nodata & $-$2.59(8) \\
J0823+0159 & 08$^{\rm h}$\;23$^{\rm m}$\;09$\fs$6 & +01$\arcdeg$\;59$\arcmin$\;12$\arcsec$ & \phm{00}4.5(\phm{0}6) & 0.82 & \phm{0}30\phm{.0}(\phm{00}5) & \phm{00}4\phm{.0}(\phm{0}2) & \phm{0}40(\phm{0}4) & \nodata & \nodata & $-$2.2\phm{0}(2) \\
J0837+0610 & 08$^{\rm h}$\;37$^{\rm m}$\;05$\fs$5 & +06$\arcdeg$\;10$\arcmin$\;17$\arcsec$ & \phm{0}12.8(\phm{0}7) & 4.55 & \phm{0}89\phm{.0}(\phm{0}14) & \phm{00}5\phm{.0}(\phm{0}1) & \phm{}766(\phm{0}9) & \phm{}286(\phm{}13) & \nodata & $-$2.4\phm{0}(2) \\
J0908$-$1739 & 09$^{\rm h}$\;08$^{\rm m}$\;38$\fs$1 & $-$17$\arcdeg$\;39$\arcmin$\;40$\arcsec$ & \phm{00}4.5(\phm{0}4) & 2.1 & \phm{0}16\phm{.0}(\phm{00}1) & \phm{00}4\phm{.0}(\phm{0}2) & \phm{0}46(\phm{0}8) & \nodata & \nodata & $-$1.6\phm{0}(1) \\
J0908$-$4913 & 09$^{\rm h}$\;08$^{\rm m}$\;35$\fs$4 & $-$49$\arcdeg$\;13$\arcmin$\;05$\arcsec$ & \phm{0}23.7(\phm{0}5) & 0.43 & \phm{0}28\phm{.0}(\phm{00}3) & \phm{0}20\phm{.0}(\phm{0}1) & \nodata & \nodata & \nodata & $-$0.30(9) \\
J0922+0638 & 09$^{\rm h}$\;22$^{\rm m}$\;14$\fs$0 & +06$\arcdeg$\;38$\arcmin$\;24$\arcsec$ & \phm{0}12.2(\phm{0}5) & 1.2 & \phm{0}52\phm{.0}(\phm{00}6) & \phm{0}10\phm{.0}(\phm{0}3) & \phm{}216(\phm{0}8) & \phm{}100(\phm{}13) & \nodata & $-$1.7\phm{0}(1) \\
J0953+0755 & 09$^{\rm h}$\;53$^{\rm m}$\;09$\fs$3 & +07$\arcdeg$\;55$\arcmin$\;37$\arcsec$ & \phm{}217.0(\phm{0}9) & 206.24 & \phm{}400\phm{.0}(\phm{}200) & \phm{}100\phm{.0}(\phm{}40) & \phm{}656(\phm{}14) & \phm{}1072(\phm{}17) & \phm{0}10.1(\phm{}2) & $-$1.1\phm{0}(5) \\
J1012$-$5857 & 10$^{\rm h}$\;12$^{\rm m}$\;48$\fs$3 & $-$58$\arcdeg$\;57$\arcmin$\;48$\arcsec$ & \phm{00}4.6(\phm{0}4) & 0.01 & \phm{0}15\phm{.0}(\phm{00}2) & \phm{00}1.9(\phm{0}1) & \nodata & \nodata & \nodata & $-$1.72(9) \\
J1041$-$1942 & 10$^{\rm h}$\;41$^{\rm m}$\;36$\fs$2 & $-$19$\arcdeg$\;42$\arcmin$\;12$\arcsec$ & \phm{00}4.5(\phm{0}4) & 0.34 & \phm{0}28\phm{.0}(\phm{00}6) & \phm{00}2.3(\phm{0}9) & \nodata & \nodata & \nodata & $-$2.2\phm{0}(3) \\
J1239+2453 & 12$^{\rm h}$\;39$^{\rm m}$\;40$\fs$2 & +24$\arcdeg$\;53$\arcmin$\;50$\arcsec$ & \phm{0}34.4(\phm{0}7) & 12.65 & \phm{}110\phm{.0}(\phm{0}33) & \phm{0}23\phm{.0}(\phm{0}5) & \phm{}136(\phm{0}6) & \nodata & \phm{00}2.7(\phm{}2) & $-$1.2\phm{0}(3) \\
J1257$-$1027 & 12$^{\rm h}$\;57$^{\rm m}$\;04$\fs$6 & $-$10$\arcdeg$\;27$\arcmin$\;05$\arcsec$ & \phm{00}2.0(\phm{0}5) & 0.17 & \phm{0}12\phm{.0}(\phm{00}1) & \phm{00}1.2(\phm{0}3) & \nodata & \nodata & \nodata & $-$1.9\phm{0}(2) \\
J1455$-$3330 & 14$^{\rm h}$\;55$^{\rm m}$\;47$\fs$9 & $-$33$\arcdeg$\;30$\arcmin$\;46$\arcsec$ & \phm{00}1.3(\phm{0}5) & 0.46 & \phm{00}9\phm{.0}(\phm{00}1) & \phm{00}0.7(\phm{0}4) & \nodata & \nodata & \nodata & $-$2.0\phm{0}(1) \\
J1532+2745 & 15$^{\rm h}$\;32$^{\rm m}$\;10$\fs$3 & +27$\arcdeg$\;45$\arcmin$\;50$\arcsec$ & \phm{00}2.9(\phm{0}4) & 1.31 & \phm{0}13\phm{.0}(\phm{00}2) & \phm{00}0.8(\phm{0}3) & \phm{0}40(\phm{0}4) & \nodata & \nodata & $-$2.0\phm{0}(2) \\
J1543$-$0620 & 15$^{\rm h}$\;43$^{\rm m}$\;30$\fs$1 & $-$06$\arcdeg$\;20$\arcmin$\;45$\arcsec$ & \phm{00}8.0(\phm{0}5) & 2.79 & \phm{0}40\phm{.0}(\phm{00}6) & \phm{00}2.0(\phm{0}7) & \phm{}369(\phm{0}4) & \phm{0}91(\phm{}12) & \phm{00}1.8(\phm{}4) & $-$2.1\phm{0}(2) \\
J1610$-$1322 & 16$^{\rm h}$\;10$^{\rm m}$\;42$\fs$7 & $-$13$\arcdeg$\;22$\arcmin$\;22$\arcsec$ & \phm{00}3.2(\phm{0}6) & 0.13 & \phm{0}16\phm{.0}(\phm{00}1) & \phm{00}1.1(\phm{0}3) & \nodata & \nodata & \nodata & $-$2.13(5) \\
J1614+0737 & 16$^{\rm h}$\;14$^{\rm m}$\;40$\fs$8 & +07$\arcdeg$\;37$\arcmin$\;33$\arcsec$ & \phm{00}1.5(\phm{0}4) & 0.3 & \phm{00}9.6(\phm{00}8) & \phm{00}0.6(\phm{0}3) & \phm{}297(\phm{}28) & \nodata & \nodata & $-$2.3\phm{0}(3) \\
J1623$-$0908 & 16$^{\rm h}$\;23$^{\rm m}$\;17$\fs$5 & $-$09$\arcdeg$\;08$\arcmin$\;49$\arcsec$ & \phm{00}1.2(\phm{0}4) & 0.03 & \phm{00}6.0(\phm{00}4) & \phm{00}0.6(\phm{0}1) & \phm{0}37(\phm{0}4) & \nodata & \nodata & $-$1.9\phm{0}(1) \\
J1703$-$1846 & 17$^{\rm h}$\;03$^{\rm m}$\;51$\fs$0 & $-$18$\arcdeg$\;46$\arcmin$\;14$\arcsec$ & \phm{00}1.9(\phm{0}4) & 0.05 & \phm{0}11\phm{.0}(\phm{00}1) & \phm{00}0.7(\phm{0}2) & \phm{}163(\phm{0}8) & \nodata & \nodata & $-$2.2\phm{0}(2) \\
J1709$-$1640 & 17$^{\rm h}$\;09$^{\rm m}$\;26$\fs$4 & $-$16$\arcdeg$\;40$\arcmin$\;57$\arcsec$ & \phm{0}17.6(\phm{0}5) & 1.93 & \phm{0}47\phm{.0}(\phm{00}5) & \phm{0}14\phm{.0}(\phm{0}3) & \phm{0}82(\phm{0}6) & \nodata & \nodata & $-$1.1\phm{0}(1) \\
J1709$-$4429 & 17$^{\rm h}$\;09$^{\rm m}$\;42$\fs$7 & $-$44$\arcdeg$\;29$\arcmin$\;07$\arcsec$ & \phm{0}16.9(\phm{0}6) & 0.11 & \phm{0}25\phm{.0}(\phm{00}4) & \phm{0}12.1(\phm{0}7) & \nodata & \nodata & \nodata & $-$0.7\phm{0}(1) \\
J1720$-$2933 & 17$^{\rm h}$\;20$^{\rm m}$\;34$\fs$1 & $-$29$\arcdeg$\;33$\arcmin$\;17$\arcsec$ & \phm{00}5.3(\phm{0}5) & 0.12 & \phm{0}32\phm{.0}(\phm{00}4) & \phm{00}1.7(\phm{0}1) & \phm{}383(\phm{}13) & \nodata & \nodata & $-$2.38(9) \\
J1722$-$3207 & 17$^{\rm h}$\;22$^{\rm m}$\;02$\fs$9 & $-$32$\arcdeg$\;07$\arcmin$\;46$\arcsec$ & \phm{0}12.2(\phm{0}5) & 0.01 & \phm{0}61\phm{.0}(\phm{00}4) & \phm{00}5.4(\phm{}11) & \phm{0}57(\phm{}12) & \phm{}229(\phm{}37) & \phm{00}1.3(\phm{}4) & $-$2.01(9) \\
J1741$-$0840 & 17$^{\rm h}$\;41$^{\rm m}$\;22$\fs$5 & $-$08$\arcdeg$\;40$\arcmin$\;31$\arcsec$ & \phm{00}5.6(\phm{0}5) & 0.09 & \phm{0}29\phm{.0}(\phm{00}8) & \phm{00}1.4(\phm{0}4) & \nodata & \nodata & \nodata & $-$2.3\phm{0}(3) \\
J1757$-$2421 & 17$^{\rm h}$\;57$^{\rm m}$\;29$\fs$3 & $-$24$\arcdeg$\;22$\arcmin$\;03$\arcsec$ & \phm{0}11.8(\phm{}13) & 0.01 & \phm{0}20\phm{.0}(\phm{00}4) & \phm{00}7.2(\phm{0}4) & \nodata & \nodata & \phm{00}2.4(\phm{}3) & $-$0.9\phm{0}(1) \\
J1759$-$2205 & 17$^{\rm h}$\;59$^{\rm m}$\;24$\fs$1 & $-$22$\arcdeg$\;05$\arcmin$\;32$\arcsec$ & \phm{00}3.8(\phm{}95) & 0.01 & \phm{0}20\phm{.0}(\phm{00}2) & \phm{00}1.3(\phm{0}1) & \phm{}139(\phm{}17) & \nodata & \nodata & $-$2.2\phm{0}(1) \\
J1807$-$0847 & 18$^{\rm h}$\;07$^{\rm m}$\;37$\fs$9 & $-$08$\arcdeg$\;47$\arcmin$\;43$\arcsec$ & \phm{0}34.9(\phm{0}9) & 0.06 & \phm{0}65\phm{.0}(\phm{00}4) & \phm{0}18\phm{.0}(\phm{0}4) & \phm{0}94(\phm{}12) & \nodata & \phm{00}4.6(\phm{}3) & $-$0.81(8) \\
J1813+4013 & 18$^{\rm h}$\;13$^{\rm m}$\;13$\fs$3 & +40$\arcdeg$\;13$\arcmin$\;39$\arcsec$ & \phm{00}2.8(\phm{0}5) & 0.18 & \phm{00}8\phm{.0}(\phm{00}2) & \phm{00}1.1(\phm{0}2) & \nodata & \nodata & \nodata & $-$1.6\phm{0}(2) \\
J1820$-$0427 & 18$^{\rm h}$\;20$^{\rm m}$\;52$\fs$5 & $-$04$\arcdeg$\;27$\arcmin$\;36$\arcsec$ & \phm{0}26.1(\phm{}11) & 0.04 & \phm{}157\phm{.0}(\phm{00}6) & \phm{0}10.1(\phm{0}2) & \phm{}975(\phm{0}8) & \phm{}499(\phm{}51) & \nodata & $-$2.18(3) \\
J1825$-$0935 & 18$^{\rm h}$\;25$^{\rm m}$\;30$\fs$6 & $-$09$\arcdeg$\;35$\arcmin$\;22$\arcsec$ & \phm{0}17.6(\phm{0}9) & 0.8 & \phm{0}36\phm{.0}(\phm{00}3) & \phm{0}10\phm{.0}(\phm{0}2) & \phm{}412(\phm{}11) & \nodata & \phm{00}0.9(\phm{}2) & $-$0.9\phm{0}(1) \\
J1829$-$1751 & 18$^{\rm h}$\;29$^{\rm m}$\;43$\fs$1 & $-$17$\arcdeg$\;51$\arcmin$\;03$\arcsec$ & \phm{0}23.6(\phm{0}6) & 0.01 & \phm{0}78\phm{.0}(\phm{00}5) & \phm{0}11\phm{.0}(\phm{0}2) & \phm{}102(\phm{0}8) & \nodata & \phm{00}3.9(\phm{}3) & $-$1.51(8) \\
J1833$-$0338 & 18$^{\rm h}$\;33$^{\rm m}$\;41$\fs$9 & $-$03$\arcdeg$\;39$\arcmin$\;02$\arcsec$ & \phm{00}9.4(\phm{0}6) & 0.01 & \phm{0}89\phm{.0}(\phm{00}5) & \phm{00}2.8(\phm{0}3) & \phm{}230(\phm{}11) & \nodata & \nodata & $-$2.79(8) \\
J1836$-$1008 & 18$^{\rm h}$\;36$^{\rm m}$\;53$\fs$9 & $-$10$\arcdeg$\;08$\arcmin$\;09$\arcsec$ & \phm{0}14.4(\phm{}11) & 0.01 & \phm{0}54\phm{.0}(\phm{00}6) & \phm{00}4.8(\phm{0}1) & \phm{0}65(\phm{}15) & \nodata & \nodata & $-$1.8\phm{0}(1) \\
J1841+0912 & 18$^{\rm h}$\;41$^{\rm m}$\;55$\fs$9 & +09$\arcdeg$\;12$\arcmin$\;08$\arcsec$ & \phm{00}4.8(\phm{0}8) & 0.28 & \phm{0}20\phm{.0}(\phm{00}1) & \phm{00}1.7(\phm{0}1) & \nodata & \nodata & \nodata & $-$1.96(6) \\
J1844+1454 & 18$^{\rm h}$\;44$^{\rm m}$\;54$\fs$8 & +14$\arcdeg$\;54$\arcmin$\;14$\arcsec$ & \phm{00}4.1(\phm{0}5) & 0.29 & \phm{0}20\phm{.0}(\phm{00}2) & \phm{00}1.8(\phm{0}4) & \phm{}105(\phm{0}7) & \nodata & \phm{00}1.7(\phm{}5) & $-$1.9\phm{0}(2) \\
J1844$-$0433 & 18$^{\rm h}$\;44$^{\rm m}$\;33$\fs$4 & $-$04$\arcdeg$\;33$\arcmin$\;12$\arcsec$ & \phm{00}3.2(\phm{0}9) & 0.01 & \phm{00}8.1(\phm{00}7) & \phm{00}1.1(\phm{0}1) & \nodata & \nodata & \nodata & $-$1.6\phm{0}(1) \\
J1847$-$0402 & 18$^{\rm h}$\;47$^{\rm m}$\;22$\fs$8 & $-$04$\arcdeg$\;02$\arcmin$\;13$\arcsec$ & \phm{0}12.6(\phm{0}7) & 0.01 & \phm{0}75\phm{.0}(\phm{00}3) & \phm{00}4.9(\phm{0}3) & \phm{}945(\phm{}14) & \nodata & \nodata & $-$2.19(6) \\
J1848$-$0123 & 18$^{\rm h}$\;48$^{\rm m}$\;23$\fs$6 & $-$01$\arcdeg$\;23$\arcmin$\;58$\arcsec$ & \phm{0}34\phm{.0}(\phm{0}3) & 0.01 & \phm{0}79\phm{.0}(\phm{00}6) & \phm{0}15\phm{.0}(\phm{0}3) & \phm{}420(\phm{}18) & \nodata & \phm{00}2.2(\phm{}3) & $-$1.2\phm{0}(1) \\
J1849$-$0636 & 18$^{\rm h}$\;49$^{\rm m}$\;06$\fs$4 & $-$06$\arcdeg$\;37$\arcmin$\;06$\arcsec$ & \phm{00}4.1(\phm{0}5) & 0.01 & \phm{0}26\phm{.0}(\phm{00}1) & \phm{00}1.4(\phm{0}1) & \phm{}203(\phm{0}8) & \nodata & \phm{00}1.1(\phm{}3) & $-$2.33(8) \\
J1850+1335 & 18$^{\rm h}$\;50$^{\rm m}$\;35$\fs$5 & +13$\arcdeg$\;35$\arcmin$\;56$\arcsec$ & \phm{00}2.3(\phm{0}4) & 0.08 & \phm{00}6\phm{.0}(\phm{00}1) & \phm{00}0.8(\phm{0}2) & \nodata & \nodata & \nodata & $-$1.5\phm{0}(2) \\
J1857+0943 & 18$^{\rm h}$\;57$^{\rm m}$\;36$\fs$3 & +09$\arcdeg$\;43$\arcmin$\;16$\arcsec$ & \phm{00}8.9(\phm{0}5) & 1.63 & \phm{0}20\phm{.0}(\phm{00}6) & \phm{00}5.0(\phm{0}5) & \nodata & \nodata & \phm{00}2.4(\phm{}2) & $-$1.2\phm{0}(2) \\
J1900$-$2600 & 19$^{\rm h}$\;00$^{\rm m}$\;47$\fs$5 & $-$26$\arcdeg$\;00$\arcmin$\;44$\arcsec$ & \phm{0}32.6(\phm{0}5) & 1.55 & \phm{}131\phm{.0}(\phm{0}12) & \phm{0}15\phm{.0}(\phm{0}3) & \phm{}408(\phm{}15) & \phm{}299(\phm{}13) & \phm{00}3.4(\phm{}3) & $-$1.7\phm{0}(1) \\
J1901+0331 & 19$^{\rm h}$\;01$^{\rm m}$\;31$\fs$8 & +03$\arcdeg$\;31$\arcmin$\;06$\arcsec$ & \phm{0}17.6(\phm{}13) & 0.01 & \phm{}165\phm{.0}(\phm{0}10) & \phm{00}4.2(\phm{0}4) & \phm{}437(\phm{}17) & \nodata & \phm{00}1.0(\phm{}3) & $-$2.89(8) \\
J1902+0556 & 19$^{\rm h}$\;02$^{\rm m}$\;42$\fs$8 & +05$\arcdeg$\;56$\arcmin$\;26$\arcsec$ & \phm{00}3.7(\phm{}96) & 0.01 & \phm{0}15\phm{.0}(\phm{00}2) & \phm{00}1.2(\phm{0}1) & \nodata & \nodata & \nodata & $-$2.0\phm{0}(1) \\
J1902+0615 & 19$^{\rm h}$\;02$^{\rm m}$\;50$\fs$3 & +06$\arcdeg$\;16$\arcmin$\;33$\arcsec$ & \phm{00}4.3(\phm{}11) & 0.01 & \phm{0}22\phm{.0}(\phm{00}4) & \phm{00}1.6(\phm{0}3) & \nodata & \nodata & \nodata & $-$2.1\phm{0}(2) \\
J1904+1011 & 19$^{\rm h}$\;04$^{\rm m}$\;02$\fs$4 & +10$\arcdeg$\;11$\arcmin$\;36$\arcsec$ & \phm{00}1.6(\phm{0}6) & 0.01 & \phm{00}4.4(\phm{00}3) & \phm{00}0.6(\phm{0}7) & \nodata & \nodata & \nodata & $-$1.6\phm{0}(1) \\
J1905$-$0056 & 19$^{\rm h}$\;05$^{\rm m}$\;27$\fs$8 & $-$00$\arcdeg$\;56$\arcmin$\;40$\arcsec$ & \phm{00}2.0(\phm{0}5) & 0.01 & \phm{00}9.8(\phm{00}6) & \phm{00}0.7(\phm{0}1) & \phm{0}38(\phm{0}7) & \nodata & \nodata & $-$2.1\phm{0}(1) \\
J1909+0254 & 19$^{\rm h}$\;09$^{\rm m}$\;38$\fs$3 & +02$\arcdeg$\;54$\arcmin$\;50$\arcsec$ & \phm{00}2.6(\phm{0}5) & 0.01 & \phm{0}21\phm{.0}(\phm{00}1) & \phm{00}0.6(\phm{0}7) & \nodata & \nodata & \nodata & $-$2.78(9) \\
J1910$-$0309 & 19$^{\rm h}$\;10$^{\rm m}$\;29$\fs$7 & $-$03$\arcdeg$\;09$\arcmin$\;54$\arcsec$ & \phm{00}2.7(\phm{0}4) & 0.01 & \phm{0}27\phm{.0}(\phm{00}3) & \phm{00}0.6(\phm{0}7) & \phm{}124(\phm{0}6) & \nodata & \nodata & $-$3.1\phm{0}(1) \\
J1913$-$0440 & 19$^{\rm h}$\;13$^{\rm m}$\;54$\fs$2 & $-$04$\arcdeg$\;40$\arcmin$\;47$\arcsec$ & \phm{0}19.1(\phm{0}4) & 0.06 & \phm{}118\phm{.0}(\phm{00}9) & \phm{00}6.8(\phm{}14) & \phm{}528(\phm{0}9) & \phm{}176(\phm{}26) & \phm{00}1.4(\phm{}3) & $-$2.3\phm{0}(1) \\
J1915+1009 & 19$^{\rm h}$\;15$^{\rm m}$\;30$\fs$0 & +10$\arcdeg$\;09$\arcmin$\;44$\arcsec$ & \phm{00}3.5(\phm{0}8) & 0.01 & \phm{0}23\phm{.0}(\phm{00}2) & \phm{00}2.0(\phm{0}4) & \nodata & \nodata & \nodata & $-$2.0\phm{0}(2) \\
J1915+1606 & 19$^{\rm h}$\;15$^{\rm m}$\;28$\fs$0 & +16$\arcdeg$\;06$\arcmin$\;30$\arcsec$ & \phm{00}1.8(\phm{0}5) & 0.01 & \phm{00}4\phm{.0}(\phm{00}1) & \phm{00}0.9(\phm{0}2) & \nodata & \nodata & \nodata & $-$1.2\phm{0}(3) \\
J1916+0951 & 19$^{\rm h}$\;16$^{\rm m}$\;32$\fs$3 & +09$\arcdeg$\;51$\arcmin$\;26$\arcsec$ & \phm{00}5.3(\phm{0}7) & 0.11 & \phm{0}20\phm{.0}(\phm{00}2) & \phm{00}1.6(\phm{0}3) & \phm{0}64(\phm{}10) & \nodata & \nodata & $-$1.9\phm{0}(1) \\
J1922+2110 & 19$^{\rm h}$\;22$^{\rm m}$\;53$\fs$4 & +21$\arcdeg$\;10$\arcmin$\;42$\arcsec$ & \phm{00}3.3(\phm{0}6) & 0.01 & \phm{0}30\phm{.0}(\phm{00}1) & \phm{00}1.4(\phm{0}2) & \phm{}131(\phm{0}9) & \nodata & \nodata & $-$2.5\phm{0}(1) \\
J1926+1648 & 19$^{\rm h}$\;26$^{\rm m}$\;45$\fs$4 & +16$\arcdeg$\;48$\arcmin$\;35$\arcsec$ & \phm{00}3.2(\phm{0}5) & 0.01 & \phm{00}8\phm{.0}(\phm{00}1) & \phm{00}1.3(\phm{0}2) & \nodata & \nodata & \nodata & $-$1.4\phm{0}(2) \\
J1932+2020 & 19$^{\rm h}$\;32$^{\rm m}$\;08$\fs$0 & +20$\arcdeg$\;20$\arcmin$\;45$\arcsec$ & \phm{00}4.5(\phm{0}6) & 0.01 & \phm{0}29\phm{.0}(\phm{00}2) & \phm{00}1.2(\phm{0}4) & \phm{}258(\phm{0}8) & \nodata & \nodata & $-$2.4\phm{0}(2) \\
J1943$-$1237 & 19$^{\rm h}$\;43$^{\rm m}$\;25$\fs$3 & $-$12$\arcdeg$\;37$\arcmin$\;41$\arcsec$ & \phm{00}2.4(\phm{0}5) & 0.21 & \phm{0}12.9(\phm{00}6) & \phm{00}1.2(\phm{0}2) & \phm{0}37(\phm{0}6) & \nodata & \nodata & $-$1.9\phm{0}(1) \\
J1949$-$2524 & 19$^{\rm h}$\;49$^{\rm m}$\;25$\fs$6 & $-$25$\arcdeg$\;23$\arcmin$\;58$\arcsec$ & \phm{00}1.3(\phm{0}4) & 0.22 & \phm{00}5.2(\phm{00}6) & \phm{00}0.4(\phm{0}1) & \nodata & \nodata & \nodata & $-$2.0\phm{0}(2) \\
J2002+3217 & 20$^{\rm h}$\;02$^{\rm m}$\;04$\fs$3 & +32$\arcdeg$\;17$\arcmin$\;18$\arcsec$ & \phm{00}2.2(\phm{0}5) & 0.02 & \phm{00}5.5(\phm{00}5) & \phm{00}1.2(\phm{0}1) & \nodata & \nodata & \nodata & $-$1.2\phm{0}(1) \\
J2006$-$0807 & 20$^{\rm h}$\;06$^{\rm m}$\;16$\fs$3 & $-$08$\arcdeg$\;07$\arcmin$\;02$\arcsec$ & \phm{00}9.7(\phm{0}4) & 1.1\phn & \phm{0}20\phm{.0}(\phm{00}3) & \phm{00}4.7(\phm{0}9) & \nodata & \nodata & \nodata & $-$1.0\phm{0}(2) \\
J2013+3845 & 20$^{\rm h}$\;13$^{\rm m}$\;10$\fs$3 & +38$\arcdeg$\;45$\arcmin$\;42$\arcsec$ & \phm{00}9.2(\phm{}11) & 0.01 & \phm{0}26\phm{.0}(\phm{00}1) & \phm{00}6.4(\phm{0}5) & \nodata & \nodata & \phm{00}2.7(\phm{}3) & $-$1.14(7) \\
J2018+2839 & 20$^{\rm h}$\;18$^{\rm m}$\;03$\fs$8 & +28$\arcdeg$\;39$\arcmin$\;54$\arcsec$ & \phm{0}53.6(\phm{}11) & 8.61 & \phm{}314\phm{.0}(\phm{0}30) & \phm{0}30\phm{.0}(\phm{}13) & \phm{}282(\phm{}10) & \nodata & \nodata & $-$2.2\phm{0}(1) \\
J2029+3744 & 20$^{\rm h}$\;29$^{\rm m}$\;23$\fs$8 & +37$\arcdeg$\;44$\arcmin$\;03$\arcsec$ & \phm{00}3.3(\phm{0}8) & 0.01 & \phm{0}18\phm{.0}(\phm{00}2) & \phm{00}0.6(\phm{0}1) & \phm{}132(\phm{}14) & \nodata & \nodata & $-$2.7\phm{0}(2) \\
J2046+1540 & 20$^{\rm h}$\;46$^{\rm m}$\;39$\fs$2 & +15$\arcdeg$\;40$\arcmin$\;32$\arcsec$ & \phm{00}3.4(\phm{0}5) & 0.13 & \phm{0}11.5(\phm{00}9) & \phm{00}1.7(\phm{0}3) & \nodata & \nodata & \nodata & $-$1.5\phm{0}(1) \\
J2046$-$0421 & 20$^{\rm h}$\;46$^{\rm m}$\;00$\fs$2 & $-$04$\arcdeg$\;21$\arcmin$\;26$\arcsec$ & \phm{00}3.5(\phm{0}6) & 0.25 & \phm{0}20\phm{.0}(\phm{00}1) & \phm{00}1.7(\phm{0}5) & \phm{0}26(\phm{0}5) & \nodata & \nodata & $-$2.1\phm{0}(2) \\
J2055+3630 & 20$^{\rm h}$\;55$^{\rm m}$\;31$\fs$4 & +36$\arcdeg$\;30$\arcmin$\;22$\arcsec$ & \phm{00}7.6(\phm{0}9) & 0.01 & \phm{0}28\phm{.0}(\phm{00}1) & \phm{00}2.6(\phm{0}1) & \phm{0}64(\phm{0}7) & \nodata & \nodata & $-$1.89(4) \\
J2124$-$3358 & 21$^{\rm h}$\;24$^{\rm m}$\;43$\fs$9 & $-$33$\arcdeg$\;58$\arcmin$\;45$\arcsec$ & \phm{00}8.2(\phm{0}5) & 5.58 & \phm{0}17\phm{.0}(\phm{00}4) & \phm{00}4.5(\phm{0}2) & \nodata & \nodata & \phm{00}2.1(\phm{}4) & $-$1.2\phm{0}(1) \\
J2129$-$5721 & 21$^{\rm h}$\;29$^{\rm m}$\;22$\fs$6 & $-$57$\arcdeg$\;21$\arcmin$\;14$\arcsec$ & \phm{00}2.4(\phm{0}3) & 0.19 & \phm{0}14\phm{.0}(\phm{00}2) & \phm{00}1.0(\phm{0}7) & \nodata & \nodata & \nodata & $-$2.1\phm{0}(1) \\
J2317+2149 & 23$^{\rm h}$\;17$^{\rm m}$\;57$\fs$9 & +21$\arcdeg$\;49$\arcmin$\;51$\arcsec$ & \phm{00}3.9(\phm{0}7) & 0.94 & \phm{0}15\phm{.0}(\phm{00}3) & \phm{00}0.9(\phm{0}5) & \nodata & \nodata & \nodata & $-$1.9\phm{0}(3) \\
\enddata
\tablenotetext{a}{Numbers quoted in parentheses are 1$-$$\sigma$ errors on the last digits of the flux densities.}
\label{tab:flux_tab}
\tablenotetext{b}{Errors in flux densities due to diffractive scintillation are quoted in addition to measurement uncertainties.}
\tablerefs{Flux densities ---  ATNF -- 400\,MHz and 1.4\,GHz \cite{psrcat}, TGSS -- 150\,MHz \cite{tgss}, MWA -- 200\,MHz \cite{MWA2017}, VLASS -- 3\,GHz \cite{vlass}}
\end{deluxetable*}

% \input{msps}
%where data came from:
% ASKAP, everything +41 dec, has polarization products on top of regular measurement, something that hasn't really been done before like this I think?
%general description of expanse of info:
% data is all below +41 dec, and taken at one specific band

%how this will improve over time:
% more area of the sky will be covered as time goes on

%Existing problems in data
%likely missing lots of low flux density Stokes V sources, detection limits stuff there 

%where things will go from here:
% data of more sky will be collected
% more accurate surveys to see lower flux density
% deeper analysis of these groups of pulsars
%what new things may be possible as improvements are made:
%

% Description of what I have done
% First, got pulsars from ATNF catalogue. 
% Then matched them with Stokes I and Stokes V objects. 
% Estimated flux density, created spectral index
% Graphed out spectral index, flux densities
% Then made cutout of position
% Looked at cutouts, took ones out that didn't match
% Calculated upper limits for ones that had one match and not another
% Did math with resulting data
% Math on resulting flux densities, polarization fraction, etc

\bibliography{main}{}
\bibliographystyle{aasjournal}

\end{document}